\documentclass[referee]{aa}  
\usepackage{graphicx}
\usepackage{natbib}

\begin{document}

\title{Hot Star Extension to the Hubble Space Telescope Stellar Spectral Library\footnote{The spectra are available in electronic form only. They are available
at the CDS (http://cdsarc.u-strasbg.fr) and MAST
(https://archive.stsci.edu).}}

\author{I. Khan\inst{\ref{inst1},\ref{inst2}}\and G. Worthey\inst{\ref{inst1},\ref{inst3}}}

\institute{Department of Physics and Astronomy, Washington State University, Pullman, WA, USA
\label{inst1}
\and
\email{islam.khan@wsu.edu}\label{inst2}
\and
\email{gworthey@wsu.edu}
\label{inst3}
}

   \date{Received 11 January, 2018; accepted 22 March, 2018}

  \abstract {Libraries of stellar spectra find many uses in
    astrophysics, from photometric calibration to stellar population
    synthesis.}  {We present low resolution spectra of 40 stars from
    0.2 $\mu$m (ultraviolet) to 1.0 $\mu$m (near infrared) with
    excellent fluxing. The stars include normal O-type stars,
    helium-burning stars, and post-asymptotic giant branch (PAGB)
    stars. } {Spectra were obtained with the Space Telescope Imaging
    Spectrograph (STIS) installed in the Hubble Space Telescope (HST)
    using three low resolution gratings, G230LB, G430L, and G750L.
    Cosmic ray hits and fringing in the red were corrected. A correction for
    scattered light was applied, significant only for our coolest
    stars. Cross-correlation was used to bring the spectra to a
    common, final, zero velocity wavelength scale. Finally, synthetic
    stellar spectra were used to estimate line of sight dust
    extinction to each star, and a five-parameter dust extinction
    model was fit, or a one-parameter fit in the case of low extinction.}  {These spectra dovetail with the similar Next
    Generation Stellar Library (NGSL) spectra, extending the NGSL's
    coverage of stellar parameters, and extending to helium burning
    stars and stars that do not fuse.}  {The fitted
    dust extinction model showed considerable variation from star to
    star, indicating variations in dust properties for different lines
    of sight. Interstellar absorption lines are present in most
    stars, notably \ion{Mg}{ii}.}

   \keywords{atlases -- stars: horizontal branch -- stars: AGB and post-AGB -- stars: early type -- ISM: dust, extinction -- Hertzsprung-Russell \& C-M diagrams}

   \titlerunning{Hot Star Spectral Library}
   \maketitle
%

\section{Introduction}

Stellar libraries are collections of stellar spectra. These libraries
find a wide variety of applications. They are used in education (e.g.,
Project CLEA and VIREO \citet{2013AAS...22125505M}), as secondary flux
standards \citep{Prugniel2001}, as checks for synthetic spectral
calculations \citep{2011MNRAS.411..807A}, as grid markers for finding
the atmospheric parameters for stars \citep{Wu2011}, as templates for
synthetic photometry \citep{2012PASP..124..140B}, in exoplanet studies
\citep{2015ApJ...804...64M}, and as ingredients in stellar population
models for integrated light \citep{1983ApJ...273..105B}. Stellar
libraries can be synthetic or observational and can differ widely in
the quality of spectrophotometric fluxing, spectral resolution, and
wavelength coverage. Empirical spectral libraries that incorporate
stars that cover a range of metal abundance include ELODIE
\citep{Prugniel2001, Prugniel2007}, CFLIB \citep{Valdes2004}, MILES
\citep{Sanchez2006}, UVES-POP \citep{2003Msngr.114...10B}, MaStar
\citep{2017arXiv170804688Y}, XSL \citep{2014A&A...565A.117C}, and IRTF
\citep{2017ApJS..230...23V}.

These libraries do not cover the ultraviolet (UV), which is of keen
interest especially for stellar population analysis of distant
galaxies. Line lists in the UV are less mature than in the optical,
and hence synthetic spectra do a worse job in predicting the flux there
\citep{2008PhST..133a4011E}. \citet{Wu1983} and \citet{Fanelli1992} presented a
UV spectral library using International Ultraviolet Explorer (IUE)
data at a resolution of 7{\AA} and a sample of 218 stars at mostly
solar abundance. 

Using the Space Telescope Imaging Spectrograph (STIS) aboard the Hubble Space Telescope (HST), the Next
Generation Spectral Library (NGSL) vastly improves spectral
resolution, sample size, and heavy element abundance coverage.
The NGSL spectra were reduced by
\citet{Heap2009jul} and made available on MAST
(http://archive.stsci.edu/prepds/stisngsl/). At the beginning of Cycle
22 the NGSL consisted of 374 stars with good coverage of temperature
and gravity and modest coverage of metallicity
\citep{Koleva2012,Heap2010}. More stars have been observed since. The
final spectral coverage is from $\sim$0.2 to $\sim$1.0$\mu$m with a
resolving power of R$\sim$1000 and roughly 3\% fluxing accuracy
\citep{Heap2009mar}.

\begin{figure}
\resizebox{\hsize}{!}{\includegraphics[width=9cm]{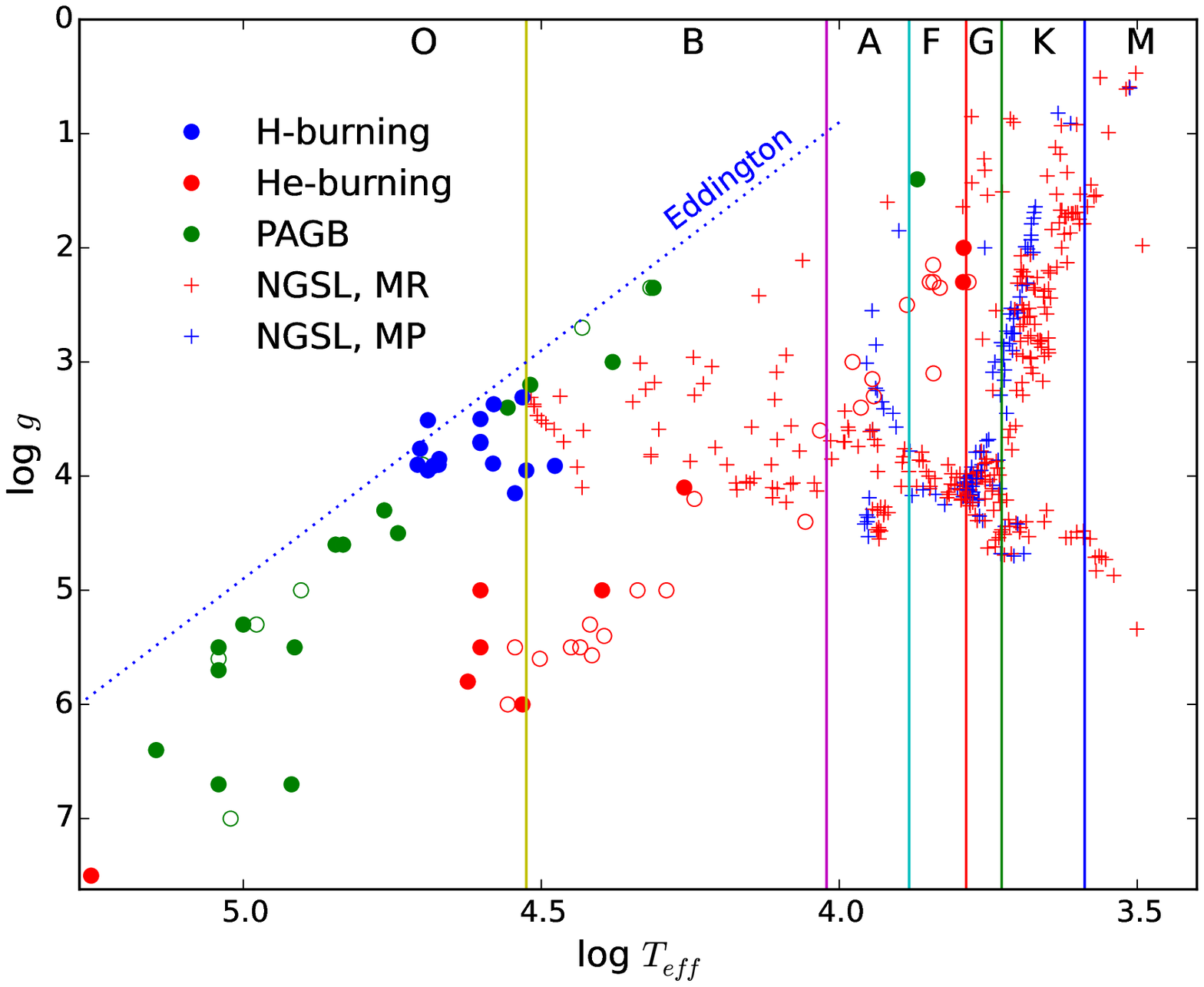}}
\caption{NGSL stars and our sample in a log $T_{{\rm eff}}$, log $g$ diagram. NGSL
  stars \citep{Koleva2012} are separated by abundance, [Fe/H] $>-1$
  (red plus) or [Fe/H] $\leq-1$ (blue plus). Hot star library
  candidate stars are marked with circles. The circles are filled for
  stars that were observed and included in this paper. All O star
  candidates (blue) were observed, but not all PAGB stars (green) or
  He-burning stars (red). To guide the eye, an approximate Eddington
  stability line is drawn, and approximate spectral type boundaries.}
\label{NGSL}
\end{figure}

The NGSL, however, is deficient in hot stars, as illustrated in
Fig. \ref{NGSL}. Normal O stars are the dominant component of the
integrated light of star forming galaxies. The NGSL also lacks
late-stage stellar types. Post-asymptotic giant
branch (PAGB) and extreme horizontal branch (EHB) stars provide most
of the UV flux for old quiescent stellar populations
\citep{1995ApJ...442..105D}. 
HST time was requested to observe these stellar groups, with the
intent to add the reduced spectra to the NGSL, broadening its scientific applicability.

The observations and data reduction, including correction for line of sight dust extinction, are described in $\S$2. 
The results for 40 (of a
total of 70 proposed) stars are given in $\S$3. Conclusions and future
work appear in $\S$4.

\section{Observations}

Stars were selected from the literature to cover a range of
temperatures in each category with a preference for brighter stars to
decrease exposure
times. \citet{1996ApJ...460..914V,Ryans2003,Mendez1988,1987MitAG..70...79H,Altmann2000,1985ApJ...299..496L,2015MNRAS.447.2404P}
were especially helpful sources. In addition, PAGB sequences in
temperature were constructed for both H-rich and H-deficient types
\citep{2003IAUS..209..169W}. SIMBAD \citep{2000A&AS..143....9W} was
used to obtain coordinates and proper motions.

The HST/STIS observations were carried out in ``snapshot mode''
(SNAP). In this mode, targets are chosen by the HST scheduling program
from a list provided by the observer, in our case, 70 objects. Over
time, targets are selected from the list and observed, but not all
targets in the list. After being scheduled, the telescope points to
the target star for somewhat less than half an orbit. After target
acquisition, spectra are taken through three low-resolution gratings
(G230LB, G430L, and G750L) whose spectral coverage overlaps at
2990-3060{\AA} and 4400-5650{\AA}. The three gratings have $\lambda /
\Delta \lambda = 1800$, 1200, and 1000, respectively. Unlike NGSL, we
specified a $0 \farcs 5$ slit instead of a $0 \farcs 2$ slit in order to
avoid a wavelength-dependent correction that varies as a function of
offset of the target from the slit center
\citep{LindlerHeap08}. Exposures are cosmic-ray split, and a fringe
flat taken for the G750L grating after each observation.

All science images were cleaned for cosmic rays with a pass through
L.A.Cosmic \citep{Dokkum2001}. Occasionally, L.A.Cosmic was
destructive to our data. In that case, we relaxed parameters and
iterated. In all cases, the L.A.Cosmic pass was followed by a second
pass through the STSDAS task ocrreject.

The STIS CCD suffers thin film interference fringing at wavelengths
redward of 7000{\AA}. The G750L fringe flats of all visits were
averaged. That average was used with STSDAS tasks \textit{normspflat},
\textit{mkfringeflat}, and \textit{defringe} to correct for the
fringing \citep{Bostroem2011}. These tasks shifted the average fringe
flat in wavelength and rescaled the correction, case by case.

\subsection{Scattered light in G230L}

Grating-internal scattered light is an issue for G230LB. It is obvious
in very cool stars. Following \citet{LindlerHeap08}, we model the
scattered light after extracting spectra from the CCD frame with a ramp plus a pedestal:
\begin{equation}
	SL = c_0(1+Slope*x),
\end{equation}

   \noindent where the scattered light $SL$ is in counts per second per pixel in the extracted net spectrum, $Slope$ is the slope of  the scattered light profile across the detector, $x$ is the pixel number, and $c_0$, the amount of scattered light, is given by:
   
\begin{eqnarray}
	c_0 = A\int{C(\lambda)/\lambda^n},
\end{eqnarray}
    
   \noindent where $C(\lambda)$ is the scattered light in counts/second and $n$ is a selectable exponent. Using $n$=3, \citet{LindlerHeap08} obtained $Slope$=0.0024 and $A$=5523 by straight line fitting.
   
The (red) scattered light calculation required integration over the
whole spectrum, not just the UV portion. We used the redder spectra to
provide these count rates, but first we converted G430L and G750L
count rates to G230LB count rates using prelaunch throughput
estimates. After assembling a full pseudo-G230L spectrum spanning the
2000 to 10000 \AA\ range, we integrated the count rate divided by the
wavelength cubed, then multiplied by 5523 to get $c_0$. We subtracted
Equation (1) from the G230L net spectrum. Thus corrected, the G230L
spectrum was multiplied by the post-launch sensitivity curve to obtain
a final spectrophotometric flux.

\subsection{Wavelength calibration}

No wavelength calibration exposures were obtained during observations
in order to save time on spacecraft overhead. Wavelength jitter is introduced by
(1) the wheel that carries the gratings (Mode Selection Mechanism)
\citep{Biretta2015}, (2) the relative radial velocity of the star and
spacecraft, and (3) pointing error along the dispersion axis. Each
exposure was cross-correlated with template stars to shift them to a
zero velocity. Templates were drawn from
banks of synthetic stellar spectra and the Calspec database \citep{2001AJ....122.2118B}, with considerable cross-checking to make sure all
templates had reliable wavelength scales. Wavelengths are in vacuum. 

\subsection{Dust extinction correction}

We adopt a five parameter model \citep{Fitzpatrick1986,
  Fitzpatrick1990} to describe the dust extinction. The model
incorporates a spectroscopic feature at $x_0 =\lambda_0^{-1}=4.595\mu
$m$^{-1}$ ($\lambda_0 \approx$ 2175 {\AA}) dubbed the extinction
  bump. The UV portion of the extinction curve 
consists of a linear ``background'' term and a Lorentzian-like
``Drude'' profile for the bump:

\begin{equation}
k(x-V)=c_1 + c_2 x + c_3 D(x;\gamma,x_0),
\end{equation}

   where $x$ is the inverse of the wavelength in $\mu$m$^{-1}$ and the
   extinction $k$ is in magnitudes. The parameters $\gamma$, $x_0$,
   and $c_3$ are bump wavelength width, center wavelength, and
   amplitude, respectively. The ``Drude'' profile is given by:
\begin{equation}
D(x;\ \gamma,x_0)=\frac{x^2}{(x^2-x^2_0)^2+x^2\gamma^2}.
\end{equation}

The parameters $c_1$ and $c_2$ follow the prescriptions  $c_2 = -0.824 + 4.717 /  R$ and $c_1 = 2.030 - 3.007 c_2$, with $c_3$ left as a free parameter.
When $R$ is allowed to vary, and for $\lambda > 2700$ {\AA} (1/$\lambda$ < 3.7 $\mu $m$^{-1}$), the
optical/IR extinction curve is constructed as a cubic spline interpolation
between a set of anchor points, following \cite{Fitzpatrick1999}. For stars with low extinction, we held all parameters constant except $A_V$ (see appendix B).

To estimate the dust extinction, dust-free templates were needed. We
generated synthetic templates using either TLUSTY
\citep{1995ApJ...439..875H} or ATLAS/SYNTHE
\citep{1993KurCD..13.....K} (and, often, both), allowing modest
temperature changes when it was clear that different templates made
for better consistency between dereddened star and template. Both
programs produced very similar continua for the same set of
atmospheric parameters. Exact matches were not attempted because our goal
was limited to fitting dust extinction parameters.

For each star/template pair, the spectra were filtered and smoothed
with cubic splines and divided to obtain an empirical extinction
curve. The empirical curve was then fit with the model described
above. Typically, five parameters, $A_v$ (absorption in magnitudes in
the V band), $R$ (ratio of total to selective extinction for diffuse
interstellar medium), $x_0$, $\gamma$, and $c_3$ were fit. For cases
of low extinction ($A_V < 0.2$ mag), $A_V$ alone was fit. The RMS
value was calculated to obtain a rough goodness of fit, but most of
the work of retrying templates and tweaking smoothing parameters was
done judiciously by eye.

\section{Results and discussion}

\begin{figure}[h]
\centering
\includegraphics[width=9.5cm]{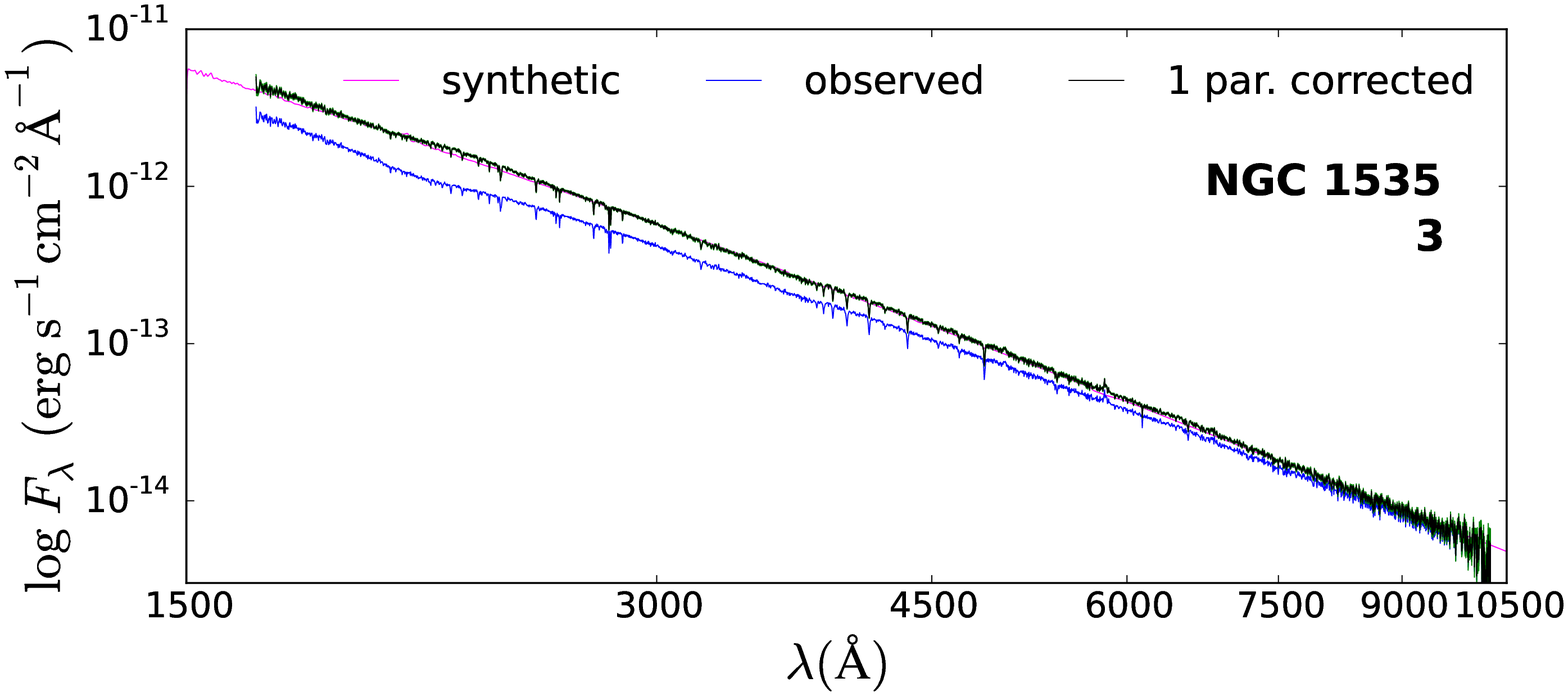}
\centering
\includegraphics[width=9.5cm]{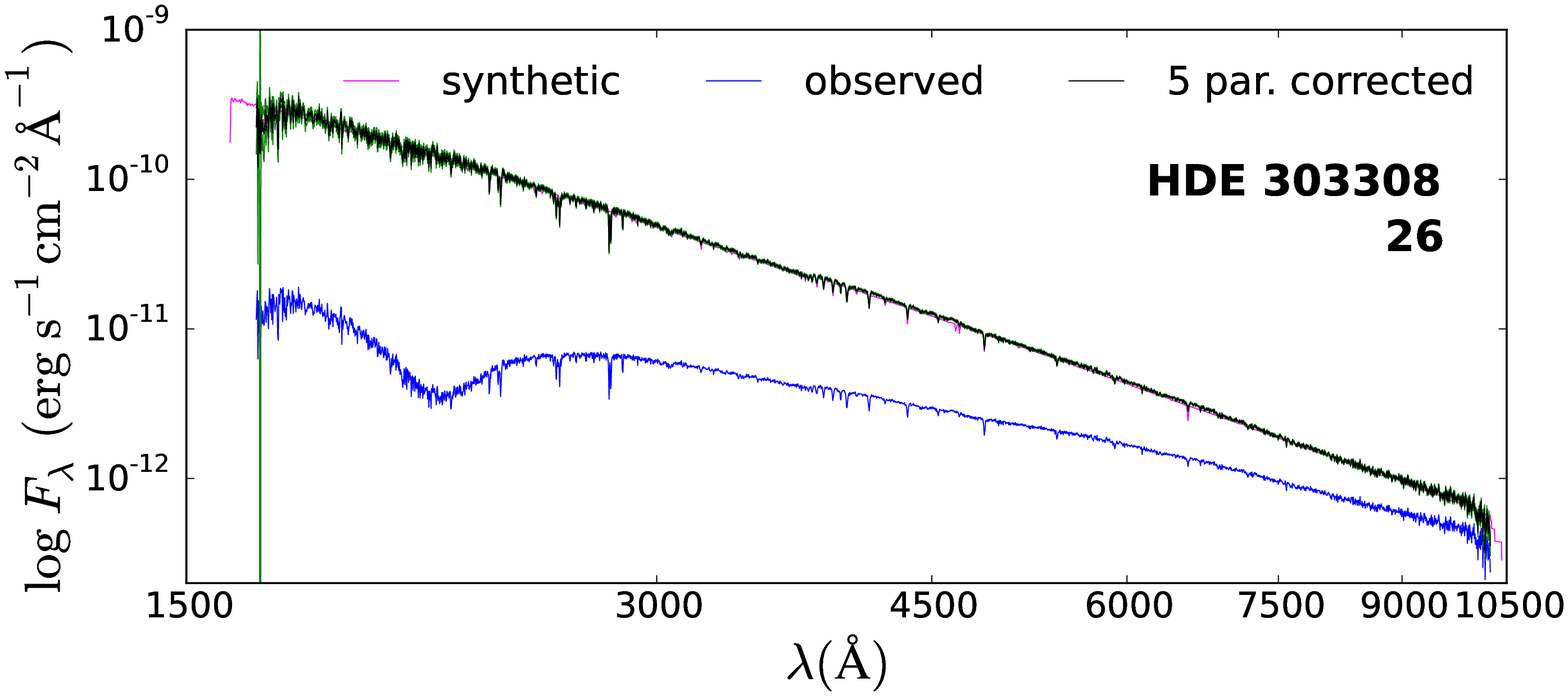}
\centering
\includegraphics[width=9.5cm]{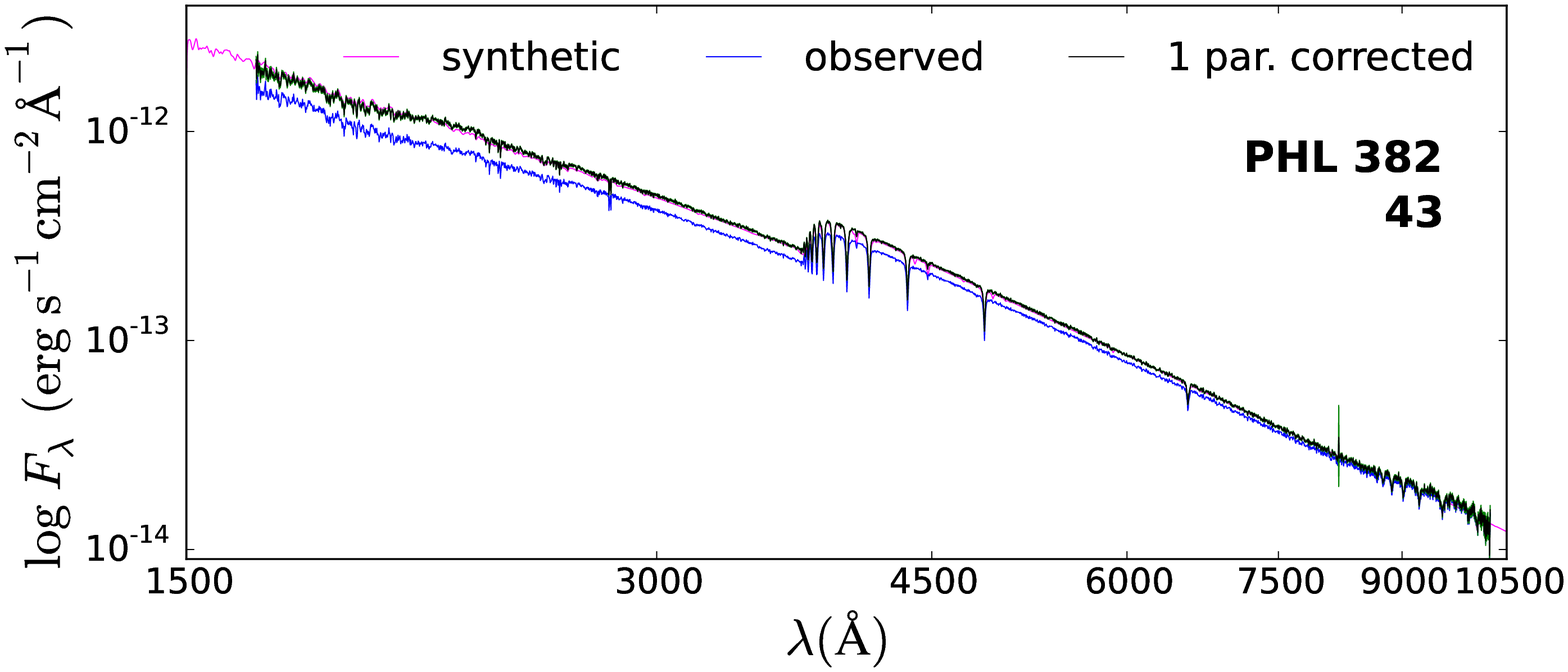}

\caption{Example spectra. Spectra as observed (blue) and as extinction corrected
  (black) with errors (flanking green curves). The no-dust template
  synthetic spectrum (magenta) overlaps. We show PAGB star \object{NGC
    1535} (top), O star \object{HDE 303308} (middle), and horizontal
  branch star \object{PHL 382} (bottom). Internal program
  identifications appear below the object name. }
\label{fig2}
\end{figure}

Our main result is to present spectra of 40 stars. The SNAP program is
still active at low priority, so a few more stars may enter the
library in the future. As best as can be contrived, the spectra are
free of cosmic rays, red fringing in the G750L, and scattered light in
the case of our coolest stars.  Cross-correlation was used to bring
the spectra to a zero velocity wavelength scale. Finally, a
five-parameter dust extinction model, or one-parameter in the case of low-extinction stars, was fit using synthetic spectra
as zero-extinction templates.

Three examples are illustrated in Fig. \ref{fig2}, and all the
stars are plotted in Appendix A. The spectra are available at the Mikulski Archive for Space Telescopes (MAST) and also at http://astro.wsu.edu/hotstarlib/. 

After the observations were reduced, we discovered a duplication. Star
3, originally filed under the name HD 26847, is the same as star 18,
NGC 1535. They arrived in the target list via different literature
paths, with temperatures 12,000 K apart. Despite this, their
separately-derived extinctions are only 0.02 mag apart, and we used
this result to gauge uncertainties arising from an uncertain
extinction law.

In the process of fitting extinction curves with synthetic templates,
the central star in \object{NGC 2392} (object 21 in our list) was
difficult to match. We computed He-rich synthetic spectra using
SYNTHE. The match improved, but was not good near H continuum
breaks. This template problem may introduce a few hundredths magnitude
error in $A_V$. 

A third of the PAGB stars show emission lines. These objects are: \object{PN A66 78} (6), \object{LS III +52 24} (13), \object{LS IV -12 111} (14), \object{IC 2448} (19), NGC 2392 (21), and \object{Hen 2-182} (22). 

A pair of resonance \ion{Mg}{ii} absorptions at 2795.5 \AA\ ($k$) and
2802.7 \AA\ ($h$) can have contributions from the stellar atmosphere
and also the interstellar medium. In stars with chromospheres, it may
also have an emission component \citep{1990PASP..102.1387G}.  As a
guide to excitation, \ion{Mg}{ii} $\lambda\lambda$4481 is weak in
early O types, but grows prominent in middle B types
\citep{2009ssc..book.....G}. Thus, in most of this library,
\ion{Mg}{ii} absorptions are interstellar (or circumstellar for
PAGB stars), not photospheric. The only unambiguously photospheric
occurrences of photospheric \ion{Mg}{ii} in the UV resonance doublet
are for cool PAGB star (13) LS III +52 24, helium-burning
stars (43) PHL 382, (47) \object{SB-707}, and (50) \object{HD 205805}, and RR Lyrae variable stars (58)
\object{DR And} and (66) \object{RX Cet}.

\section{Conclusions}

\indent We present low resolution spectra from 0.2 to 1.0$\mu$m, fully
fluxed and wavelength calibrated, observed with STIS on HST and
extending into the near UV. This newly-observed collection of spectra
adds diagnostic value to the NGSL by exploring O stars, helium burning
stars, and Post-AGB stars. Dealing with dust extinction proved to be
essential for most stars, and we fit corrections in the form of a
polynomial in 1/$\lambda$ with a ``Drude'' absorption profile to
account for the UV bump.

Future work includes making all spectra observed with the same STIS
configuration over the history of HST available for general use. Honed
stellar atmospheric parameters is another goal of the project. The
library can be used for improving synthetic spectra, observational
flux calibration, and even basic teaching about spectral types and
stellar atmospheres. Extensions to this project should include fitting
the library and making stellar population models for integrated light
similar to those of
\citet{2016MNRAS.463.3409V}. At that point, the library will be
applied to galaxy population synthesis. Due to its UV coverage, it
will be uniquely applicable to high redshift galaxies. Due to its
inclusion of very hot stars, it will be applicable to starburst
objects both local and cosmological.

\begin{acknowledgements}
      
Support for this work was provided by NASA through grant number
HST-GO-14141 from the Space Telescope Science Institute, which is
operated by AURA, Inc., under NASA contract NAS 5-26555. Based on
observations made with the NASA/ESA Hubble Space Telescope, obtained
from the data archive at the Space Telescope Science Institute. STScI
is operated by the Association of Universities for Research in
Astronomy, Inc. under NASA contract NAS 5-26555.

\end{acknowledgements}

\bibliography{bib}

\begin{thebibliography}{}
\expandafter\ifx\csname natexlab\endcsname\relax\def\natexlab#1{#1}\fi

\bibitem[{{Allende Prieto}(2011)}]{2011MNRAS.411..807A}
{Allende Prieto}, C. 2011, \mnras, 411, 807

\bibitem[{{Altmann} \& {de Boer}(2000)}]{Altmann2000}
{Altmann}, M., \& {de Boer}, K.~S. 2000, \aap, 353, 135

\bibitem[{{Bagnulo} {et~al.}(2003){Bagnulo}, {Jehin}, {Ledoux}, {Cabanac},
  {Melo}, {Gilmozzi}, \& {ESO Paranal Science Operations
  Team}}]{2003Msngr.114...10B}
{Bagnulo}, S., {Jehin}, E., {Ledoux}, C., {et~al.} 2003, The Messenger, 114, 10

\bibitem[{{Bessell} \& {Murphy}(2012)}]{2012PASP..124..140B}
{Bessell}, M., \& {Murphy}, S. 2012, \pasp, 124, 140

\bibitem[{{Biretta} {et~al.}(2015){Biretta}, {a}, {a}, {a}, {a}, \&
  {a}}]{Biretta2015}
{Biretta}, J., {a}, A., {a}, A., {et~al.} 2015

\bibitem[{{Bohlin} {et~al.}(2001){Bohlin}, {Dickinson}, \&
  {Calzetti}}]{2001AJ....122.2118B}
{Bohlin}, R.~C., {Dickinson}, M.~E., \& {Calzetti}, D. 2001, \aj, 122, 2118

\bibitem[{{Bostroem} \& {Proffitt}(2011)}]{Bostroem2011}
{Bostroem}, K., \& {Proffitt}, C. 2011

\bibitem[{{Bruzual A.}(1983)}]{1983ApJ...273..105B}
{Bruzual A.}, G. 1983, \apj, 273, 105

\bibitem[{{Chen} {et~al.}(2014){Chen}, {Trager}, {Peletier}, {Lan{\c c}on},
  {Vazdekis}, {Prugniel}, {Silva}, \& {Gonneau}}]{2014A&A...565A.117C}
{Chen}, Y.-P., {Trager}, S.~C., {Peletier}, R.~F., {et~al.} 2014, \aap, 565,
  A117

\bibitem[{{Dorman} {et~al.}(1995){Dorman}, {O'Connell}, \&
  {Rood}}]{1995ApJ...442..105D}
{Dorman}, B., {O'Connell}, R.~W., \& {Rood}, R.~T. 1995, \apj, 442, 105

\bibitem[{{Edvardsson}(2008)}]{2008PhST..133a4011E}
{Edvardsson}, B. 2008, Physica Scripta Volume T, 133, 014011

\bibitem[{{Fanelli} {et~al.}(1992){Fanelli}, {O'Connell}, {Burstein}, \&
  {Wu}}]{Fanelli1992}
{Fanelli}, M.~N., {O'Connell}, R.~W., {Burstein}, D., \& {Wu}, C.-C. 1992,
  \apjs, 82, 197

\bibitem[{Fitzpatrick(1999)}]{Fitzpatrick1999}
Fitzpatrick, E.~L. 1999, Publications of the Astronomical Society of the
  Pacific, 111, 63

\bibitem[{{Fitzpatrick} \& {Massa}(1986)}]{Fitzpatrick1986}
{Fitzpatrick}, E.~L., \& {Massa}, D. 1986, \apj, 307, 286

\bibitem[{{Fitzpatrick} \& {Massa}(1990)}]{Fitzpatrick1990}
---. 1990, \apjs, 72, 163

\bibitem[{{Gray} \& {Corbally}(2009)}]{2009ssc..book.....G}
{Gray}, R.~O., \& {Corbally}, J., C. 2009, {Stellar Spectral Classification}
  (Princeton University Press)

\bibitem[{{Gurzadian} {et~al.}(1990){Gurzadian}, {Cholakian}, {Kondo}, {Shore},
  \& {Terzian}}]{1990PASP..102.1387G}
{Gurzadian}, G.~A., {Cholakian}, V.~G., {Kondo}, Y., {Shore}, S.~N., \&
  {Terzian}, Y. 1990, PASP, 102, 1387

\bibitem[{{Heap}(2009)}]{Heap2009jul}
{Heap}, S. 2009, {Constraining the Star-Formation and Metal-Enrichment
  Histories of Galaxies with the Next Generation Spectral Library}, HST
  Proposal

\bibitem[{{Heap} \& {Lindler}(2009)}]{Heap2009mar}
{Heap}, S., \& {Lindler}, D.~J. 2009, Astrophysics and Space Science
  Proceedings, 7, 273

\bibitem[{{Heap} \& {Lindler}(2010)}]{Heap2010}
{Heap}, S.~R., \& {Lindler}, D. 2010, in Bulletin of the American Astronomical
  Society, Vol.~42, American Astronomical Society Meeting Abstracts \#215, 494

\bibitem[{{Heber}(1987)}]{1987MitAG..70...79H}
{Heber}, U. 1987, Mitteilungen der Astronomischen Gesellschaft Hamburg, 70, 79

\bibitem[{{Heber} {et~al.}(1984){Heber}, {Hamann}, {Hunger}, {Kudritzki},
  {Simon}, \& {Mendez}}]{Heber1984}
{Heber}, U., {Hamann}, W.-R., {Hunger}, K., {et~al.} 1984, \aap, 136, 331

\bibitem[{{Hubeny} \& {Lanz}(1995)}]{1995ApJ...439..875H}
{Hubeny}, I., \& {Lanz}, T. 1995, \apj, 439, 875

\bibitem[{{Koleva, M.} \& {Vazdekis, A.}(2012)}]{Koleva2012}
{Koleva, M.}, \& {Vazdekis, A.} 2012, \aap, 538, A143

\bibitem[{{Kurucz}(1993)}]{1993KurCD..13.....K}
{Kurucz}, R. 1993, ATLAS9 Stellar Atmosphere Programs and 2 km/s grid.~Kurucz
  CD-ROM No.~13.~ Cambridge, Mass.: Smithsonian Astrophysical Observatory,
  1993., 13

\bibitem[{{Lamontagne} {et~al.}(1985){Lamontagne}, {Wesemael}, {Fontaine}, \&
  {Sion}}]{1985ApJ...299..496L}
{Lamontagne}, R., {Wesemael}, F., {Fontaine}, G., \& {Sion}, E.~M. 1985, \apj,
  299, 496

\bibitem[{{Lindler} \& {Heap}(2008)}]{LindlerHeap08}
{Lindler}, D., \& {Heap}, S.~R. 2008, {STIS Next Generation Spectral Library}

\bibitem[{{Lynas-Gray} {et~al.}(1984){Lynas-Gray}, {Schoenberner}, {Hill}, \&
  {Heber}}]{1984MNRAS.209..387L}
{Lynas-Gray}, A.~E., {Schoenberner}, D., {Hill}, P.~W., \& {Heber}, U. 1984,
  \mnras, 209, 387

\bibitem[{{Mann} {et~al.}(2015){Mann}, {Feiden}, {Gaidos}, {Boyajian}, \& {von
  Braun}}]{2015ApJ...804...64M}
{Mann}, A.~W., {Feiden}, G.~A., {Gaidos}, E., {Boyajian}, T., \& {von Braun},
  K. 2015, \apj, 804, 64

\bibitem[{{Marschall} {et~al.}(2013){Marschall}, {Snyder}, \&
  {Cooper}}]{2013AAS...22125505M}
{Marschall}, L.~A., {Snyder}, G., \& {Cooper}, P. 2013, in American
  Astronomical Society Meeting Abstracts, Vol. 221, American Astronomical
  Society Meeting Abstracts \#221, 255.05

\bibitem[{{Mendez} {et~al.}(1988){Mendez}, {Kudritzki}, {Herrero}, {Husfeld},
  \& {Groth}}]{Mendez1988}
{Mendez}, R.~H., {Kudritzki}, R.~P., {Herrero}, A., {Husfeld}, D., \& {Groth},
  H.~G. 1988, \aap, 190, 113

\bibitem[{{Pancino} {et~al.}(2015){Pancino}, {Britavskiy}, {Romano},
  {Cacciari}, {Mucciarelli}, \& {Clementini}}]{2015MNRAS.447.2404P}
{Pancino}, E., {Britavskiy}, N., {Romano}, D., {et~al.} 2015, \mnras, 447, 2404

\bibitem[{{Pereira} {et~al.}(2012){Pereira}, {Gallino}, \& {Bisterzo}}]{refId0}
{Pereira}, C.~B., {Gallino}, R., \& {Bisterzo}, S. 2012, \aap, 538, A48

\bibitem[{{Prugniel} \& {Soubiran}(2001)}]{Prugniel2001}
{Prugniel}, P., \& {Soubiran}, C. 2001, \aap, 369, 1048

\bibitem[{{Prugniel} {et~al.}(2007){Prugniel}, {Soubiran}, {Koleva}, \& {Le
  Borgne}}]{Prugniel2007}
{Prugniel}, P., {Soubiran}, C., {Koleva}, M., \& {Le Borgne}, D. 2007, ArXiv
  Astrophysics e-prints, astro-ph/0703658

\bibitem[{{Ryans} {et~al.}(2003){Ryans}, {Dufton}, {Mooney}, {Rolleston},
  {Keenan}, {Hubeny}, \& {Lanz}}]{Ryans2003}
{Ryans}, R.~S.~I., {Dufton}, P.~L., {Mooney}, C.~J., {et~al.} 2003, \aap, 401,
  1119

\bibitem[{{S{\'a}nchez-Bl{\'a}zquez} {et~al.}(2006){S{\'a}nchez-Bl{\'a}zquez},
  {Peletier}, {Jim{\'e}nez-Vicente}, {Cardiel}, {Cenarro},
  {Falc{\'o}n-Barroso}, {Gorgas}, {Selam}, \& {Vazdekis}}]{Sanchez2006}
{S{\'a}nchez-Bl{\'a}zquez}, P., {Peletier}, R.~F., {Jim{\'e}nez-Vicente}, J.,
  {et~al.} 2006, \mnras, 371, 703

\bibitem[{Sarkar {et~al.}(2012)Sarkar, García-Hernández, Parthasarathy,
  Manchado, García-Lario, \& Takeda}]{Sarker2012}
Sarkar, G., García-Hernández, D.~A., Parthasarathy, M., {et~al.} 2012,
  Monthly Notices of the Royal Astronomical Society, 421, 679

\bibitem[{{Vacca} {et~al.}(1996){Vacca}, {Garmany}, \&
  {Shull}}]{1996ApJ...460..914V}
{Vacca}, W.~D., {Garmany}, C.~D., \& {Shull}, J.~M. 1996, \apj, 460, 914

\bibitem[{{Valdes} {et~al.}(2004){Valdes}, {Gupta}, {Rose}, {Singh}, \&
  {Bell}}]{Valdes2004}
{Valdes}, F., {Gupta}, R., {Rose}, J.~A., {Singh}, H.~P., \& {Bell}, D.~J.
  2004, \apjs, 152, 251

\bibitem[{{van Dokkum}(2001)}]{Dokkum2001}
{van Dokkum}, P.~G. 2001, \pasp, 113, 1420

\bibitem[{{Vazdekis} {et~al.}(2016){Vazdekis}, {Koleva}, {Ricciardelli},
  {R{\"o}ck}, \& {Falc{\'o}n-Barroso}}]{2016MNRAS.463.3409V}
{Vazdekis}, A., {Koleva}, M., {Ricciardelli}, E., {R{\"o}ck}, B., \&
  {Falc{\'o}n-Barroso}, J. 2016, \mnras, 463, 3409

\bibitem[{{Villaume} {et~al.}(2017){Villaume}, {Conroy}, {Johnson}, {Rayner},
  {Mann}, \& {van Dokkum}}]{2017ApJS..230...23V}
{Villaume}, A., {Conroy}, C., {Johnson}, B., {et~al.} 2017, \apjs, 230, 23

\bibitem[{{Wenger} {et~al.}(2000){Wenger}, {Ochsenbein}, {Egret}, {Dubois},
  {Bonnarel}, {Borde}, {Genova}, {Jasniewicz}, {Lalo{\"e}}, {Lesteven}, \&
  {Monier}}]{2000A&AS..143....9W}
{Wenger}, M., {Ochsenbein}, F., {Egret}, D., {et~al.} 2000, \aaps, 143, 9

\bibitem[{{Werner} {et~al.}(2003){Werner}, {Deetjen}, {Dreizler}, {Rauch}, \&
  {Kruk}}]{2003IAUS..209..169W}
{Werner}, K., {Deetjen}, J.~L., {Dreizler}, S., {Rauch}, T., \& {Kruk}, J.~W.
  2003, 209, 169

\bibitem[{{Werner} \& {Herwig}(2006)}]{2006PASP..118..183W}
{Werner}, K., \& {Herwig}, F. 2006, \pasp, 118, 183

\bibitem[{{Wu} {et~al.}(1983){Wu}, {Ake}, {Boggess}, {Bohlin}, {Imhoff},
  {Holm}, {Levay}, {Panek}, {Schiffer}, \& {Turnrose}}]{Wu1983}
{Wu}, C.-C., {Ake}, T.~B., {Boggess}, A., {et~al.} 1983, NASA IUE Newsl.,
  No.~22, 2+324 pp., 22

\bibitem[{{Wu} {et~al.}(2011){Wu}, {Singh}, {Prugniel}, {Gupta}, \&
  {Koleva}}]{Wu2011}
{Wu}, Y., {Singh}, H.~P., {Prugniel}, P., {Gupta}, R., \& {Koleva}, M. 2011,
  \aap, 525, A71

\bibitem[{{Yan} \& {the MaStar Team}(2017)}]{2017arXiv170804688Y}
{Yan}, R., \& {the MaStar Team}. 2017, ArXiv e-prints, arXiv:1708.04688

\end{thebibliography}

\appendix

\section{Gallery of spectra}

\begin{figure*}[!htb]
\minipage{0.49\textwidth}
  \includegraphics[width=\linewidth]{fig_s03.eps}
\endminipage\hfill
\minipage{0.49\textwidth}
  \includegraphics[width=\linewidth]{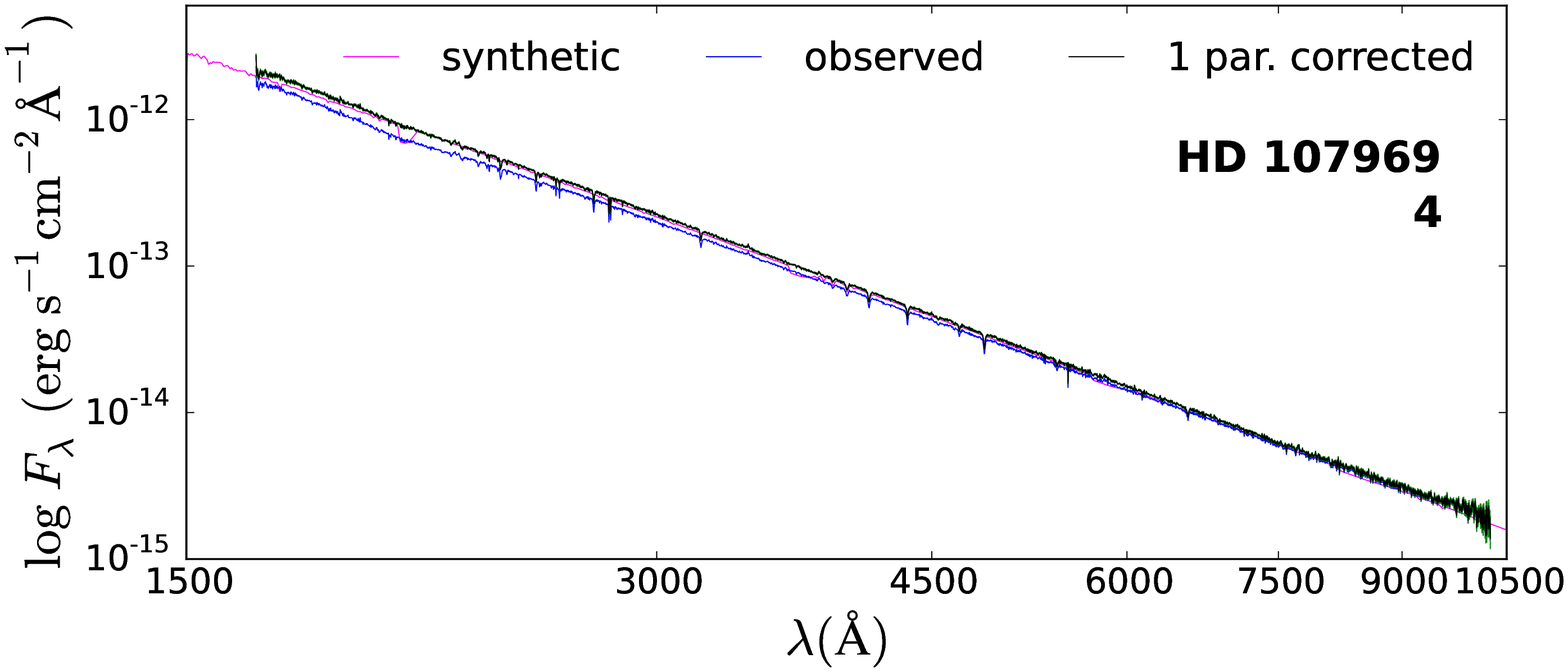} 
\endminipage\hfill
\minipage{0.49\textwidth}
  \includegraphics[width=\linewidth]{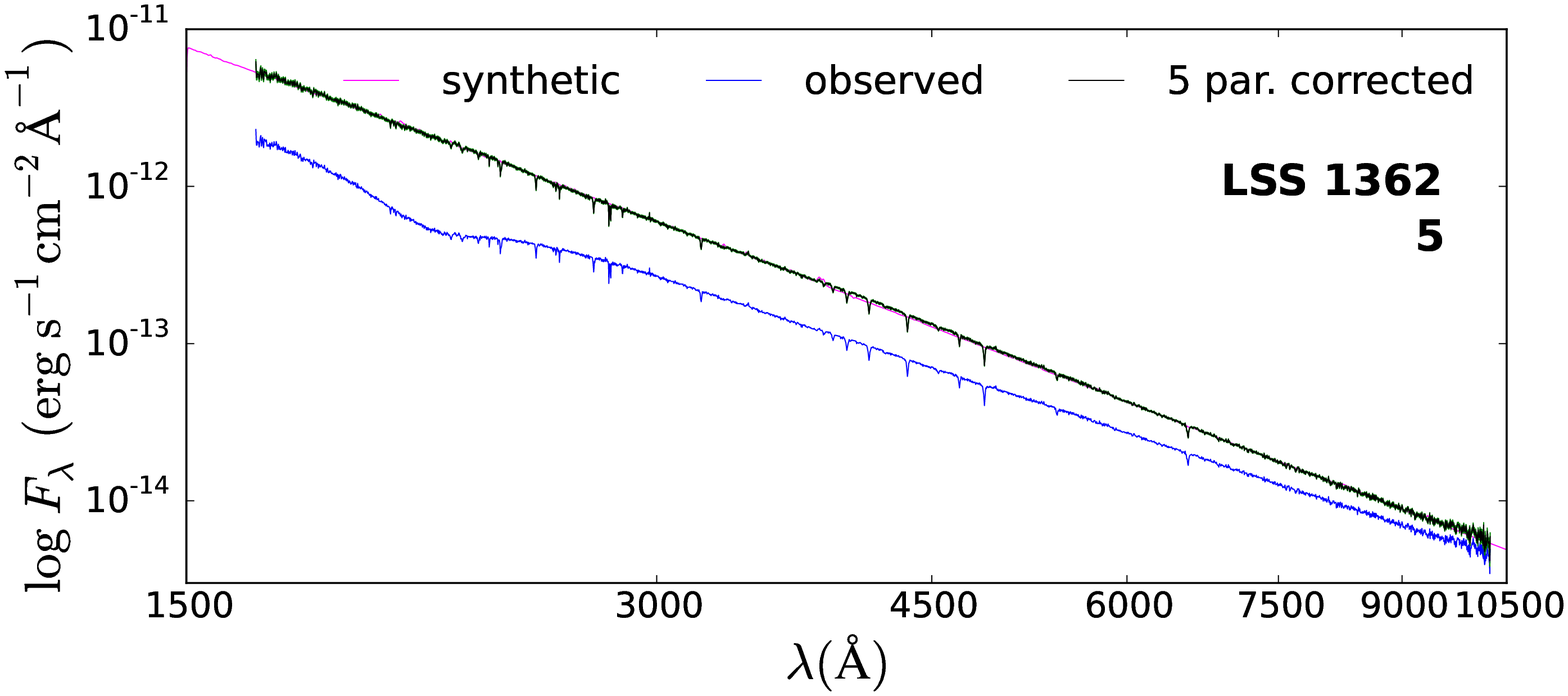}
\endminipage\hfill
\minipage{0.49\textwidth}
  \includegraphics[width=\linewidth]{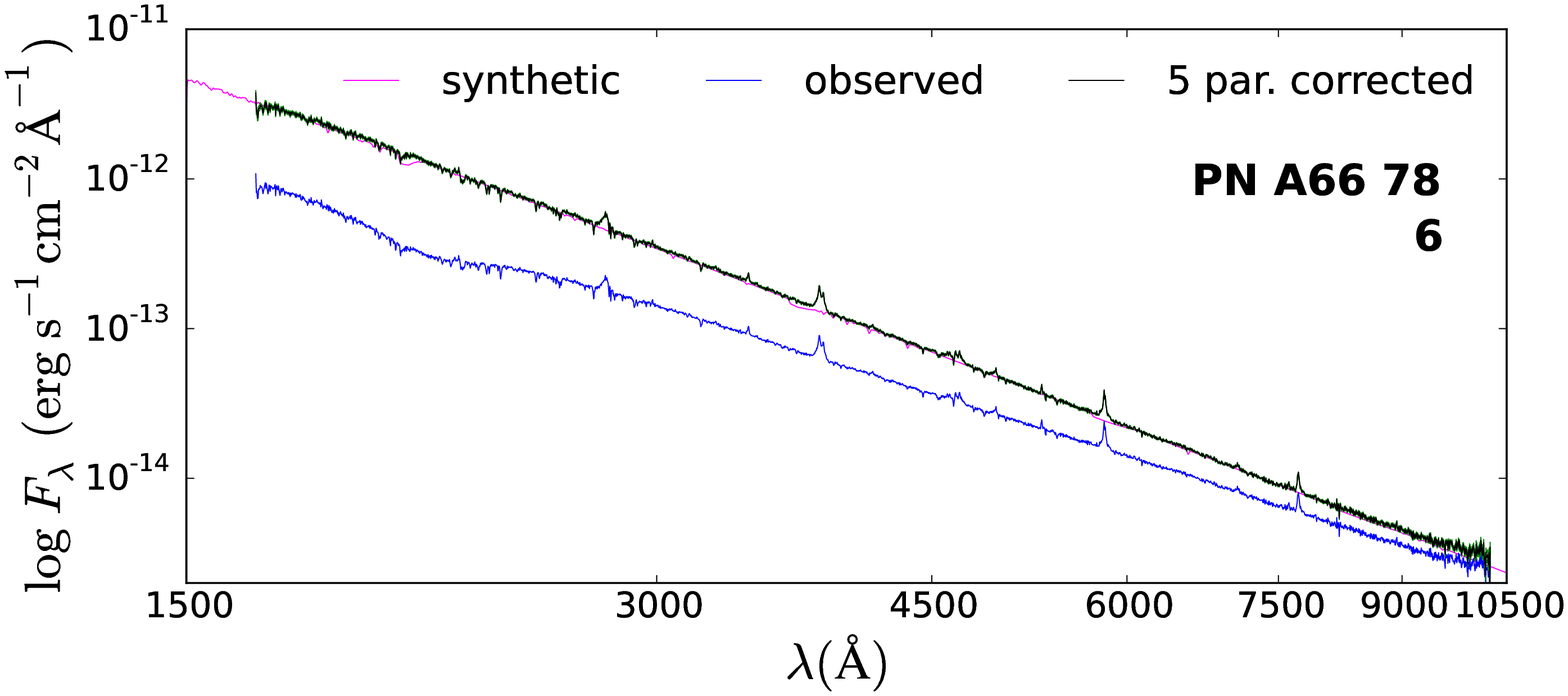}
\endminipage\hfill
\minipage{0.49\textwidth}
  \includegraphics[width=\linewidth]{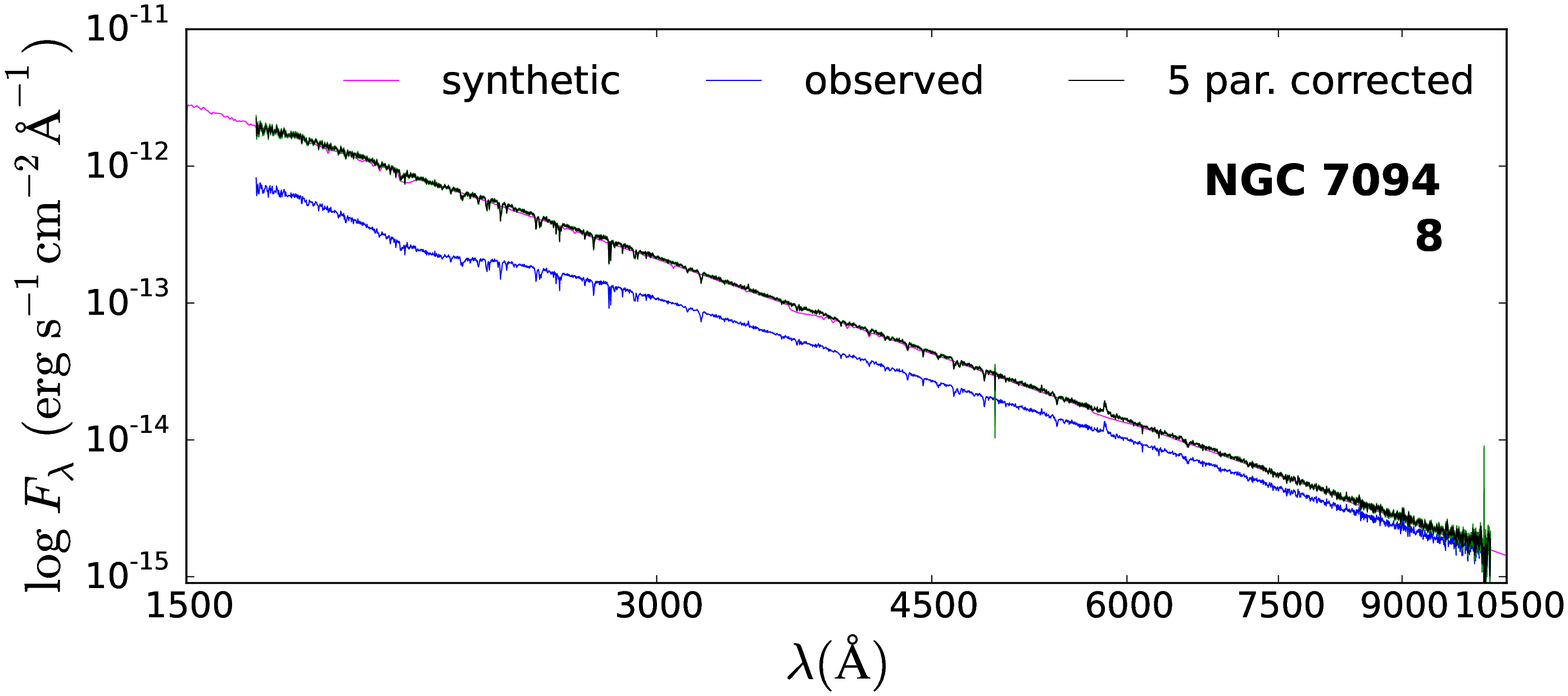}
\endminipage\hfill
\minipage{0.49\textwidth}
  \includegraphics[width=\linewidth]{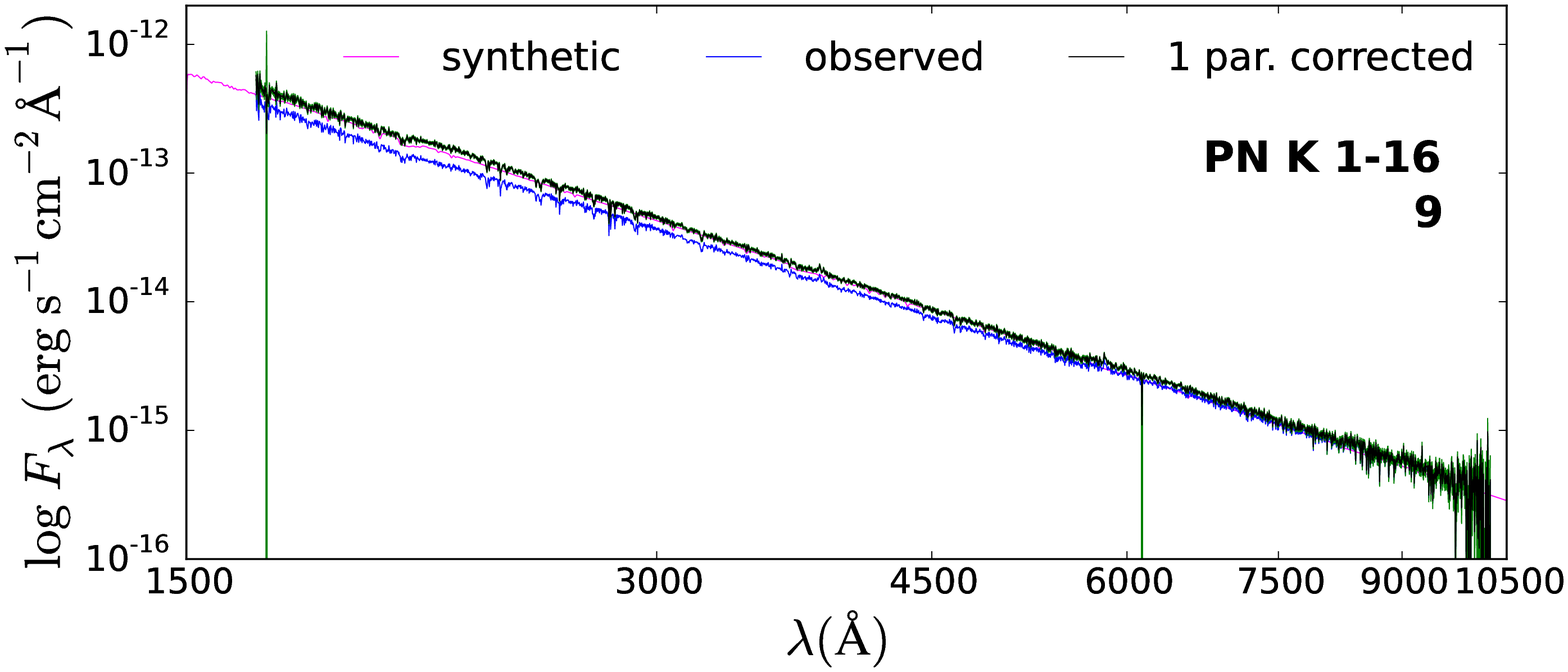} 
\endminipage\hfill
\minipage{0.49\textwidth}
  \includegraphics[width=\linewidth]{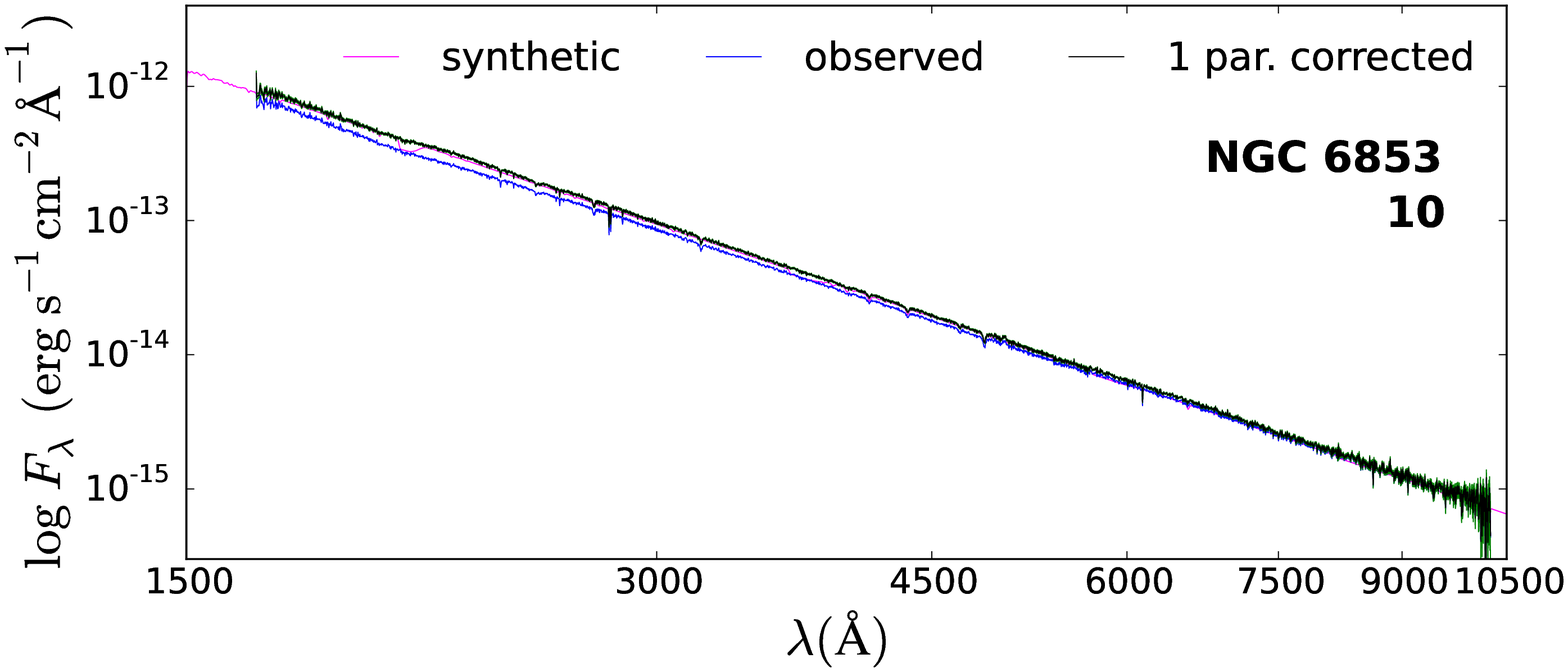}
\endminipage\hfill
\minipage{0.49\textwidth}
  \includegraphics[width=\linewidth]{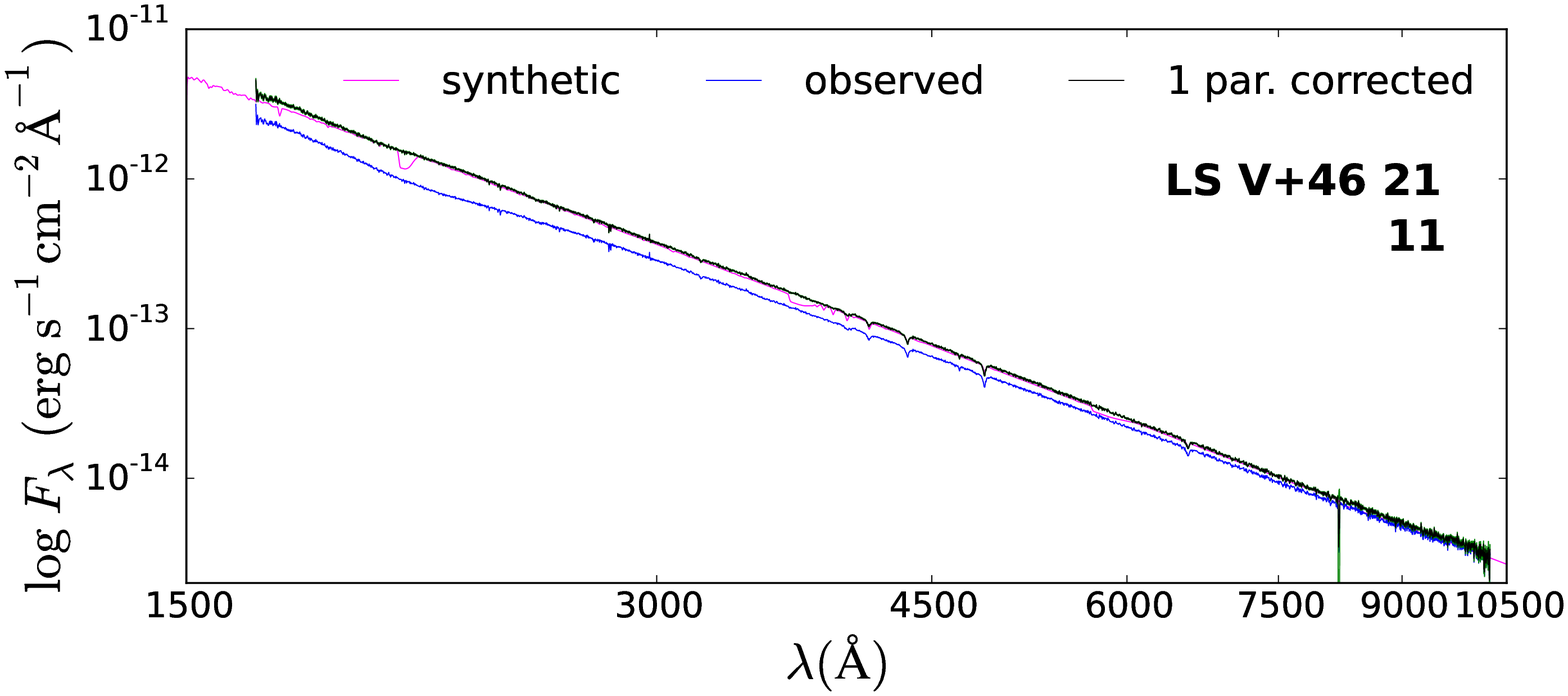}
\endminipage\hfill
\minipage{0.49\textwidth}
  \includegraphics[width=\linewidth]{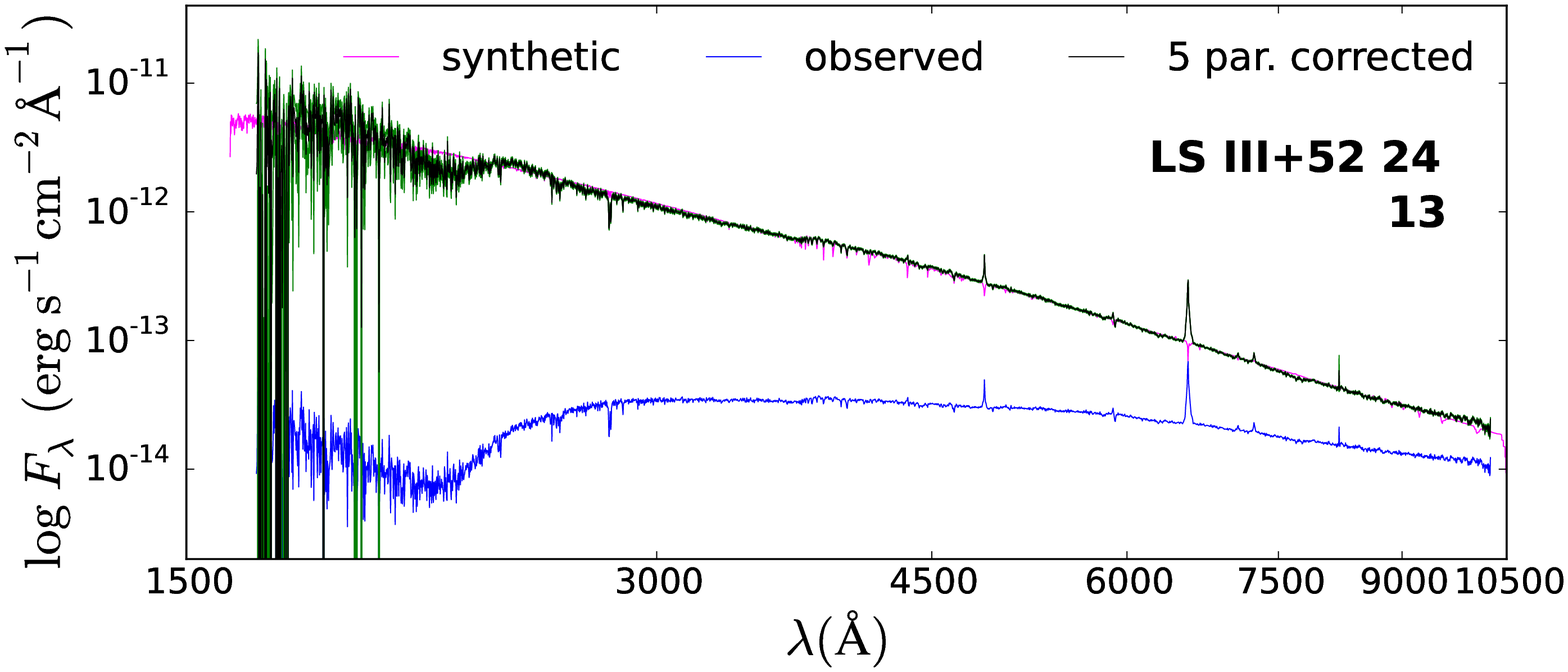} 
\endminipage\hfill
\minipage{0.49\textwidth}
  \includegraphics[width=\linewidth]{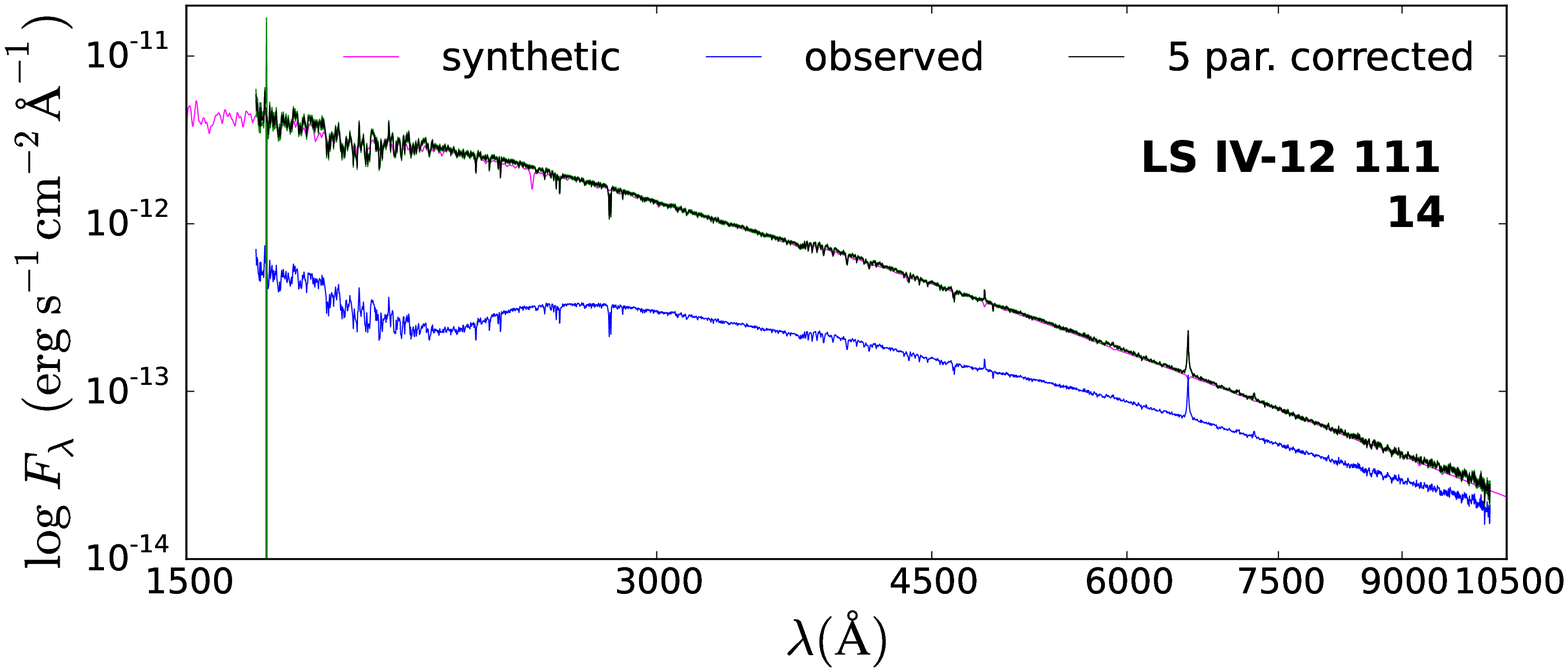}
\endminipage\hfill
\minipage{0.49\textwidth}
  \includegraphics[width=\linewidth]{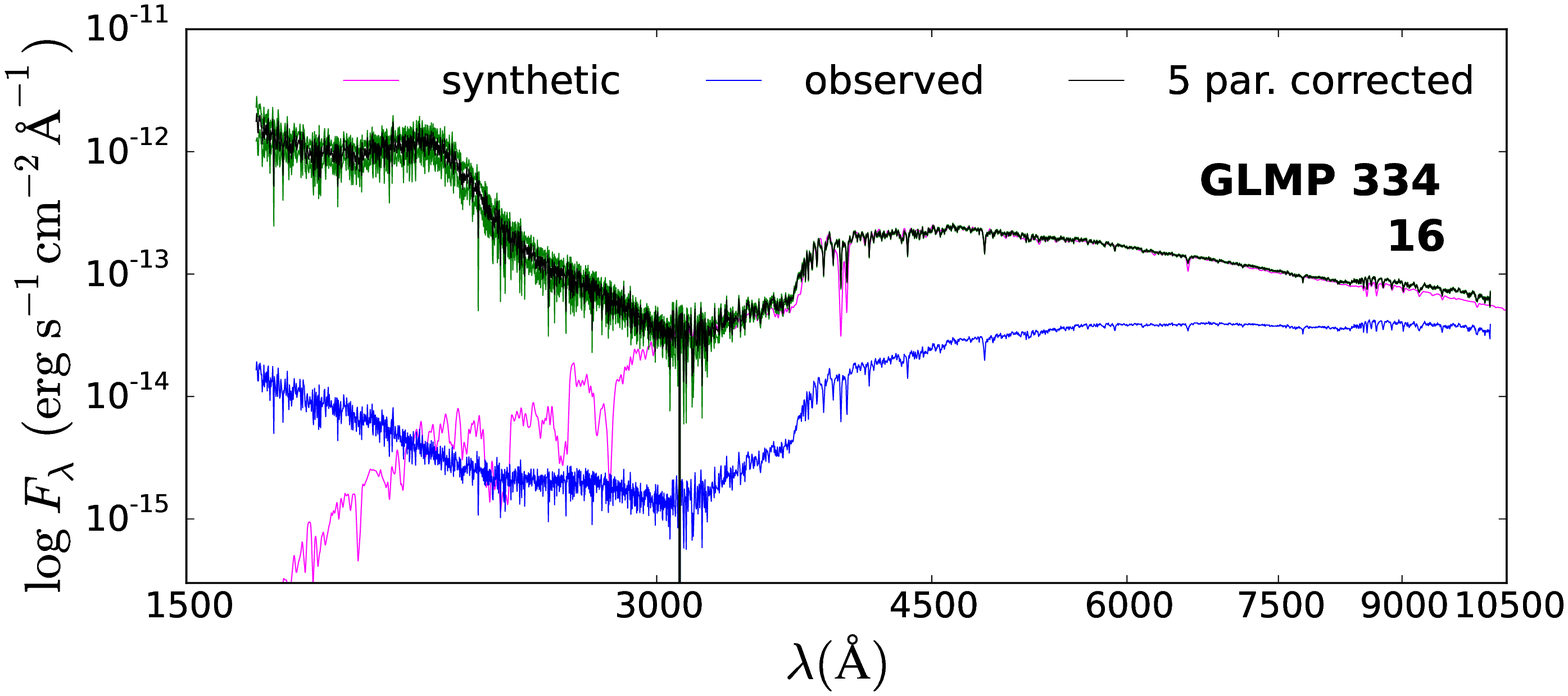}
\endminipage\hfill
\minipage{0.49\textwidth}
  \includegraphics[width=\linewidth]{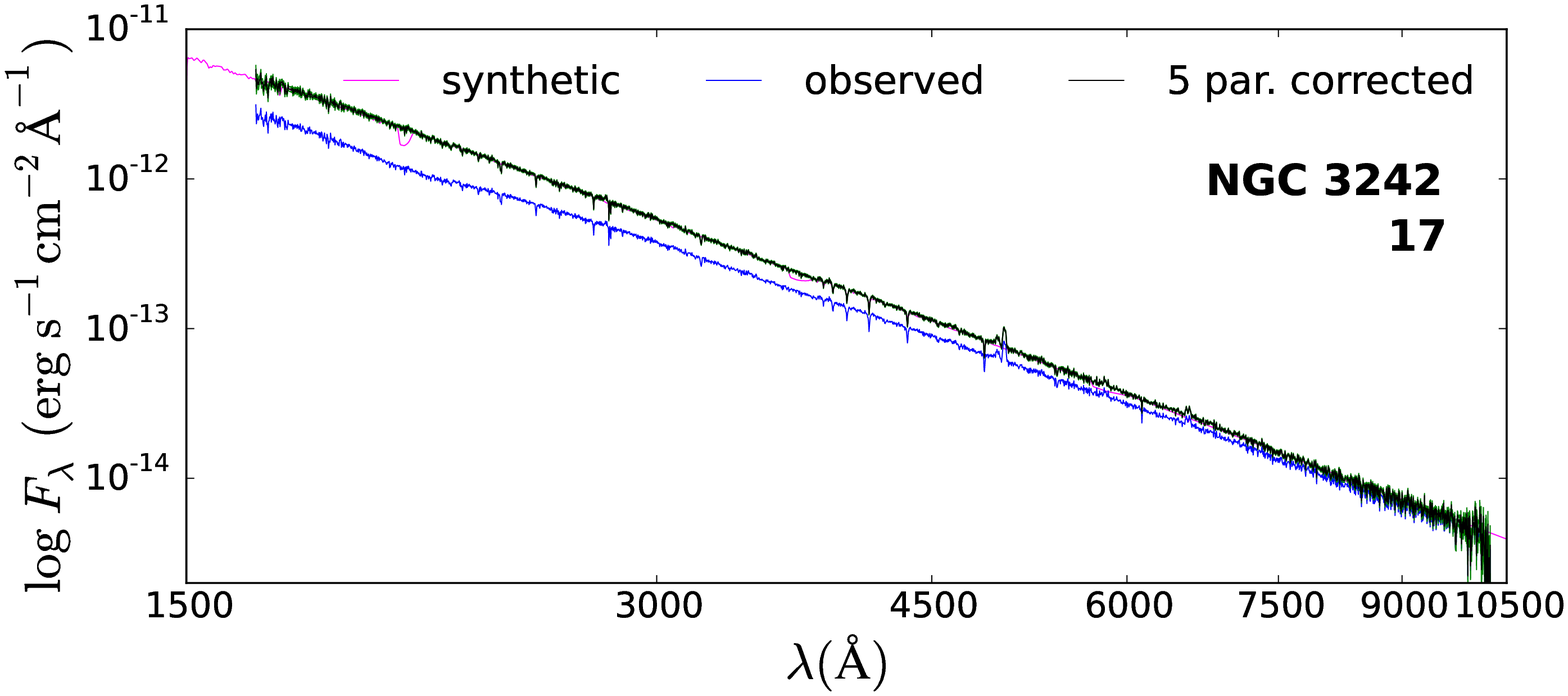}
\endminipage\hfill
\minipage{0.49\textwidth}
  \includegraphics[width=\linewidth]{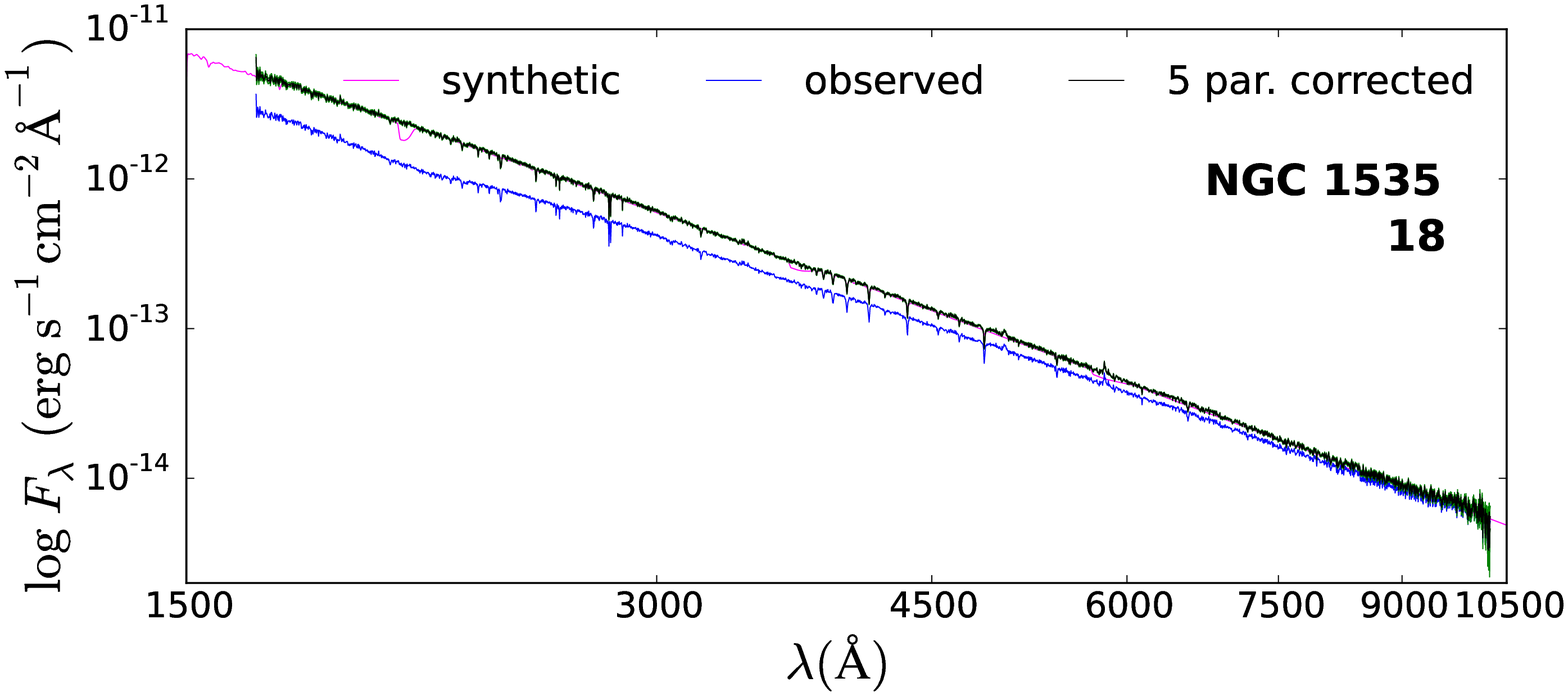} 
\endminipage\hfill
\minipage{0.49\textwidth}
  \includegraphics[width=\linewidth]{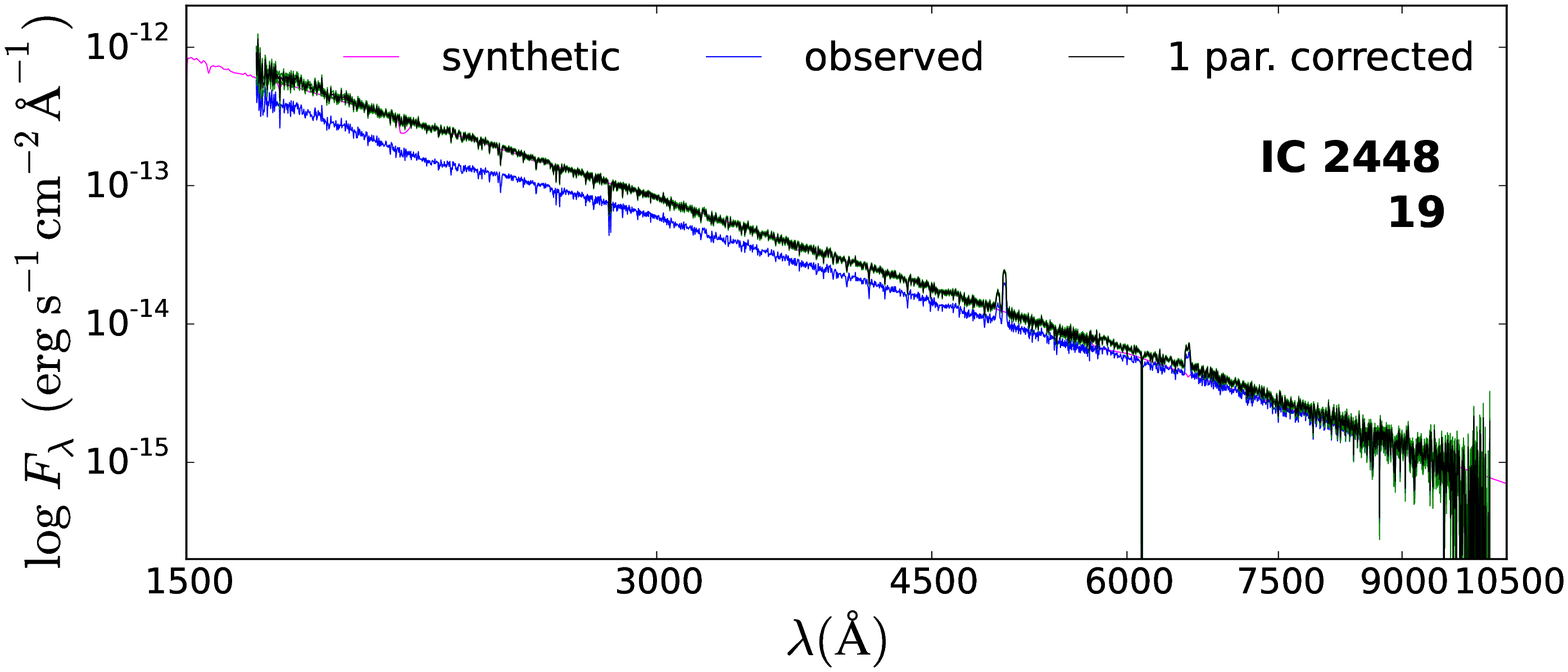}
\endminipage\hfill
\caption{Library spectra. Fluxes (blue) and extinction-corrected fluxes (black with
  green error bounds) are plotted along with the corresponding
  synthetic spectral template (magenta). }
\end{figure*}

\begin{figure*}[!htb]
\minipage{0.49\textwidth}
  \includegraphics[width=\linewidth]{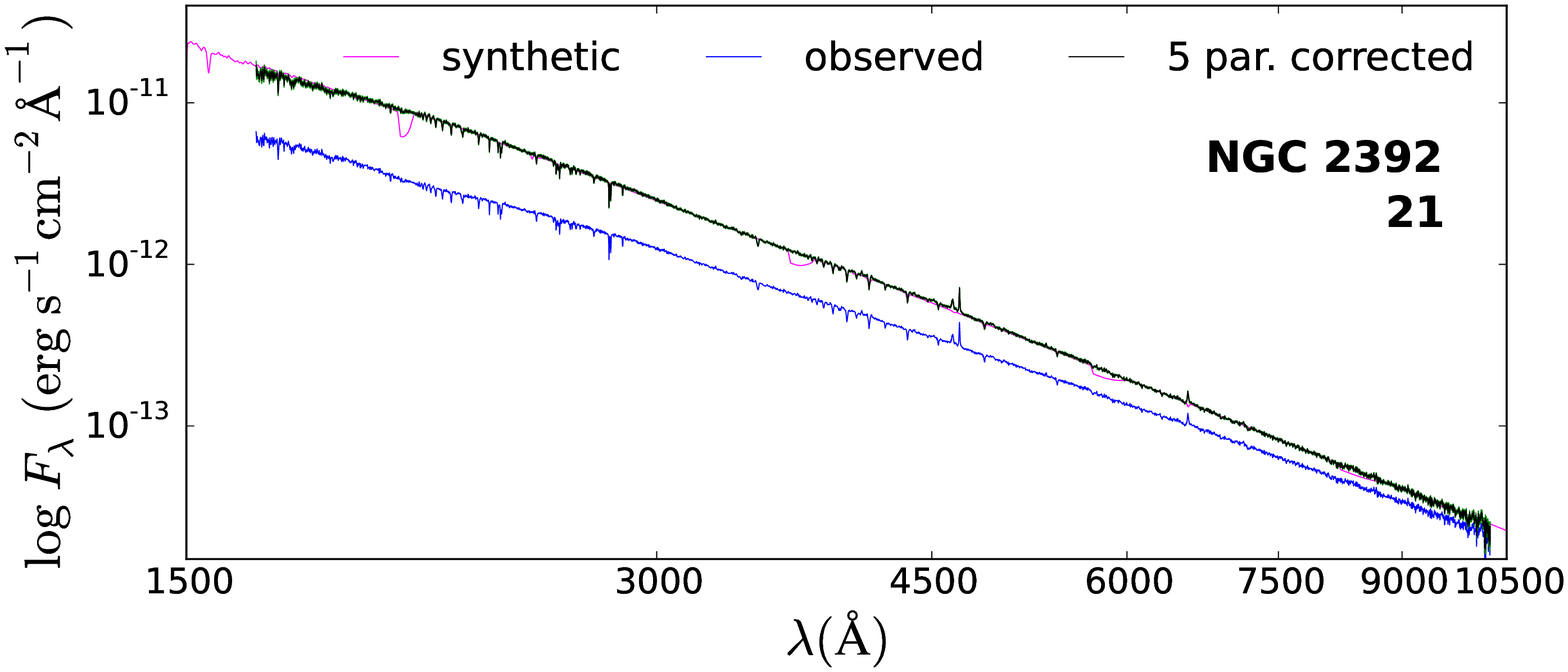}
\endminipage\hfill
\minipage{0.49\textwidth}
  \includegraphics[width=\linewidth]{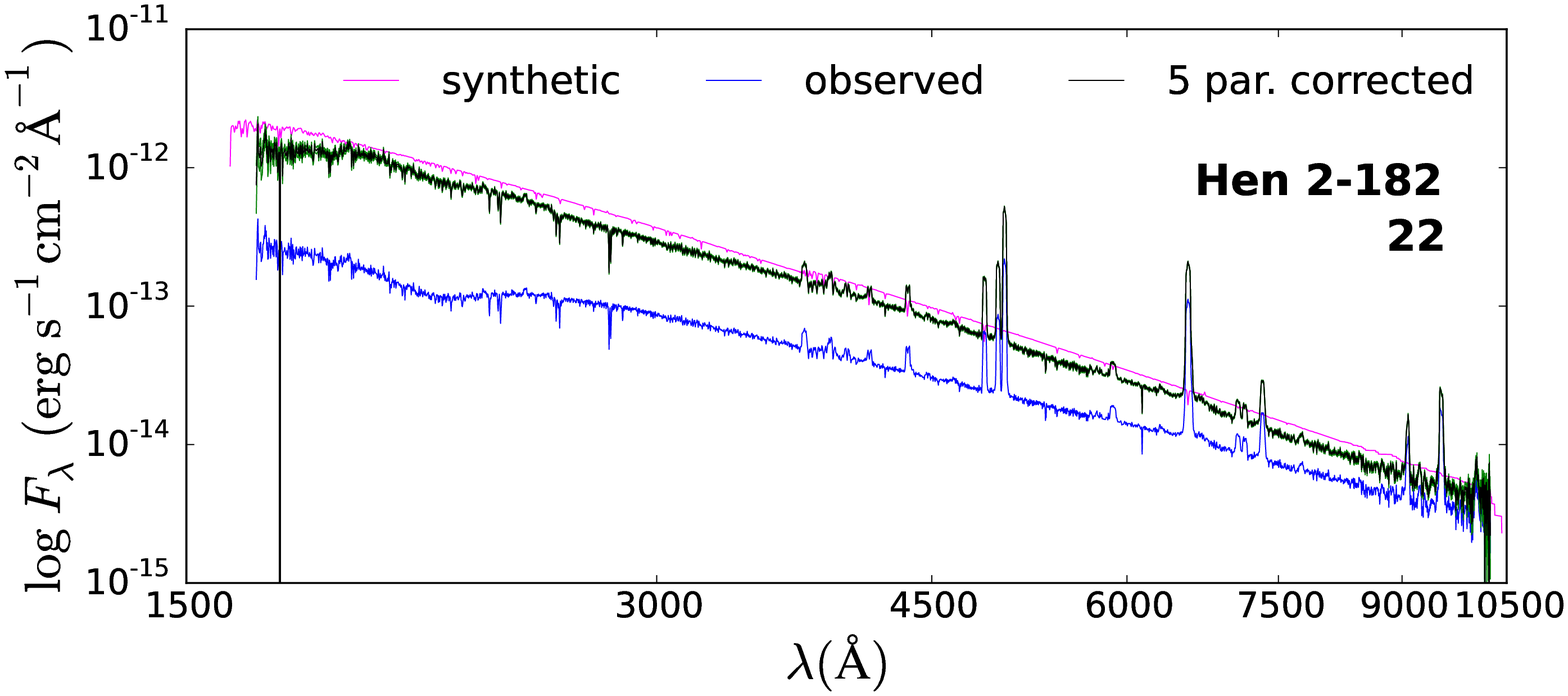}
\endminipage\hfill
\minipage{0.49\textwidth}
  \includegraphics[width=\linewidth]{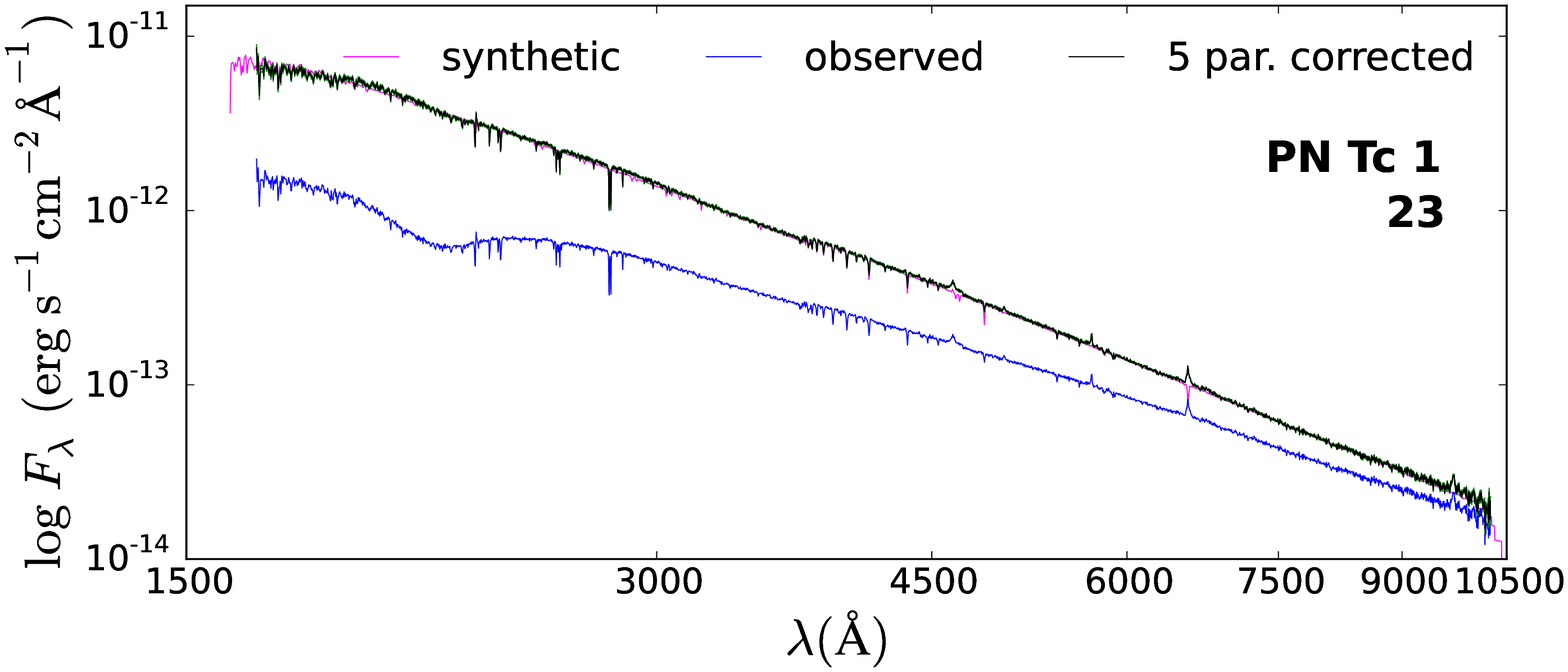}
\endminipage\hfill
\minipage{0.49\textwidth}
  \includegraphics[width=\linewidth]{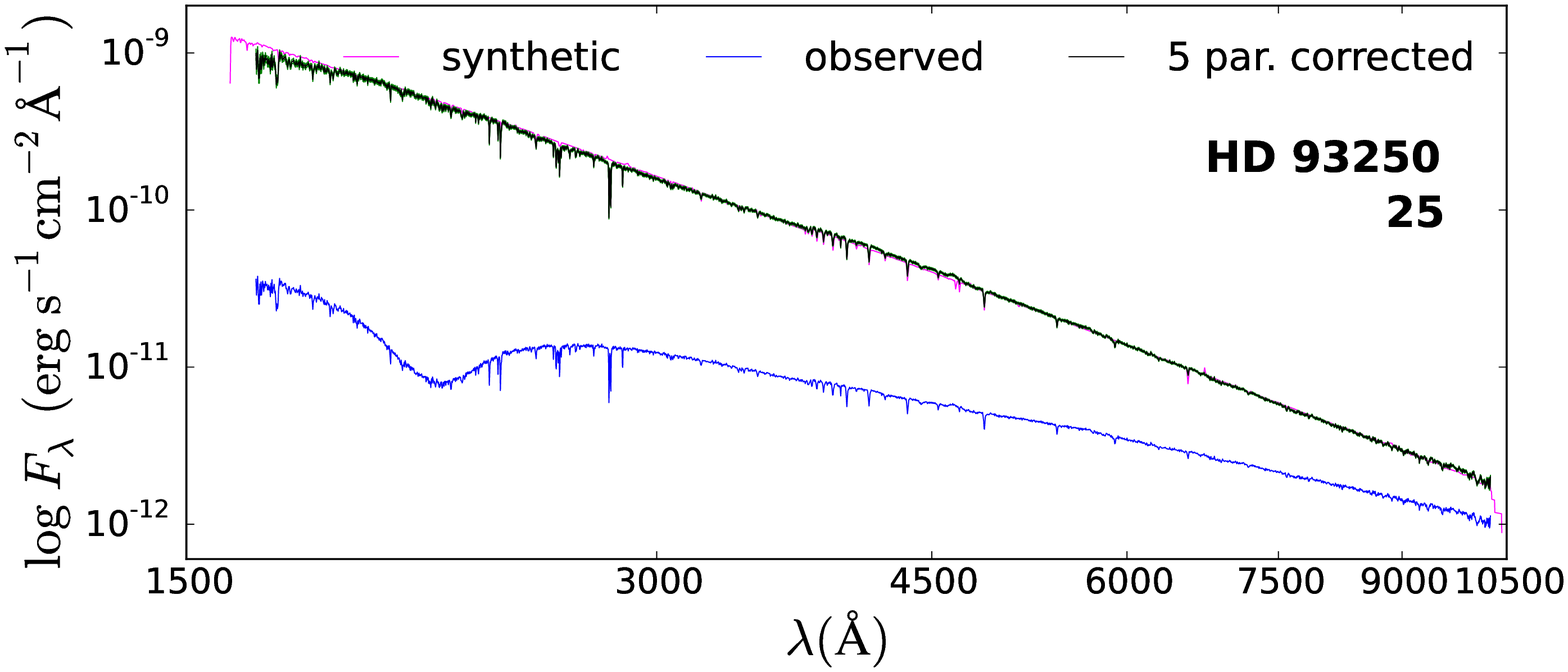}
\endminipage\hfill
\minipage{0.49\textwidth}
  \includegraphics[width=\linewidth]{fig_s26.eps} 
\endminipage\hfill
\minipage{0.49\textwidth}
  \includegraphics[width=\linewidth]{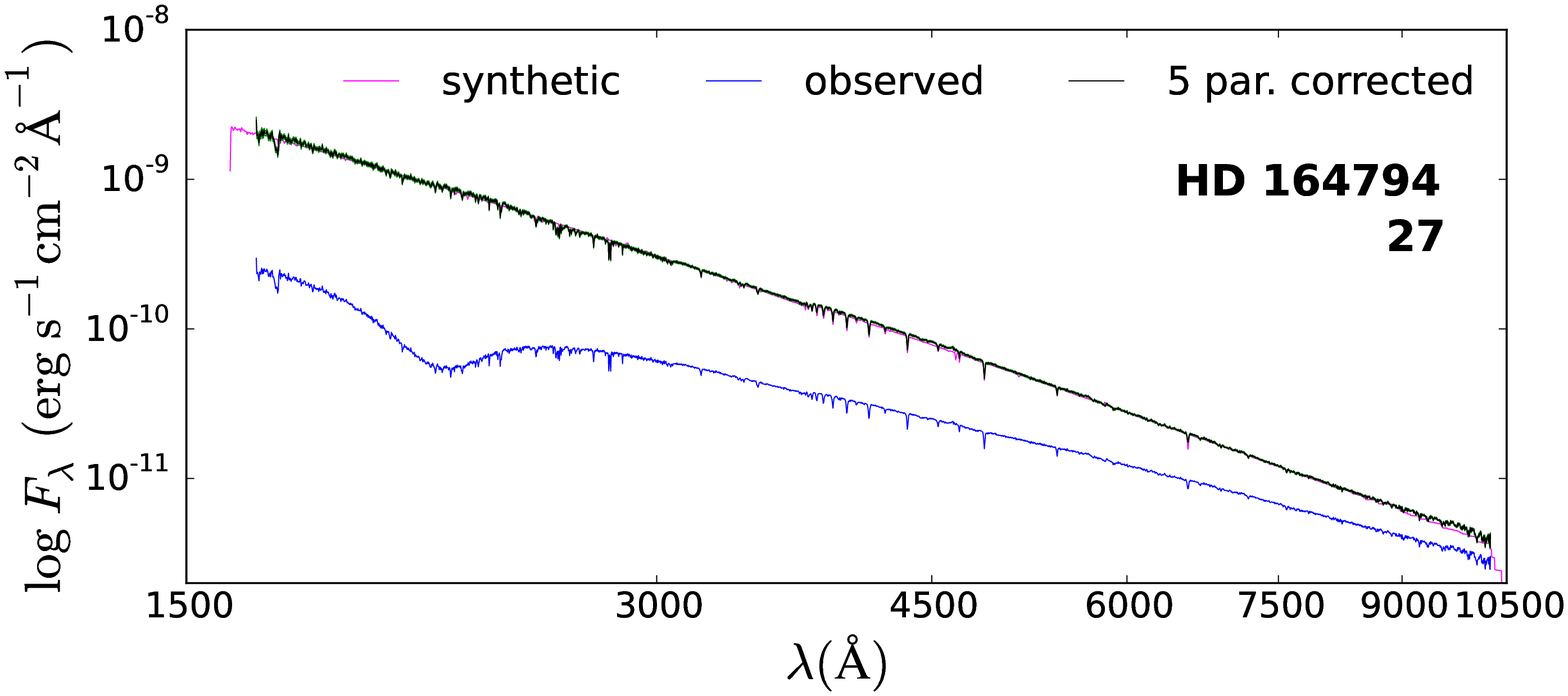}
\endminipage\hfill
\minipage{0.49\textwidth}
  \includegraphics[width=\linewidth]{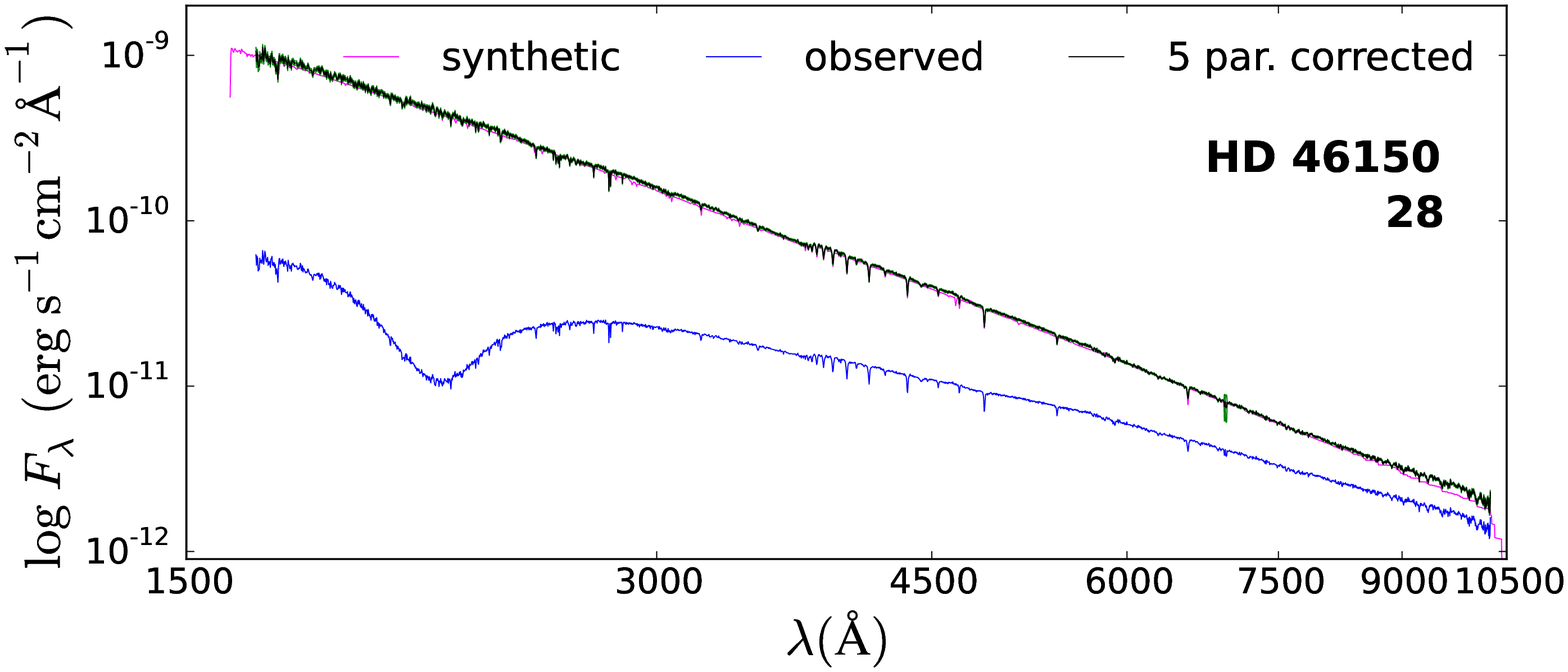}
\endminipage\hfill
\minipage{0.49\textwidth}
  \includegraphics[width=\linewidth]{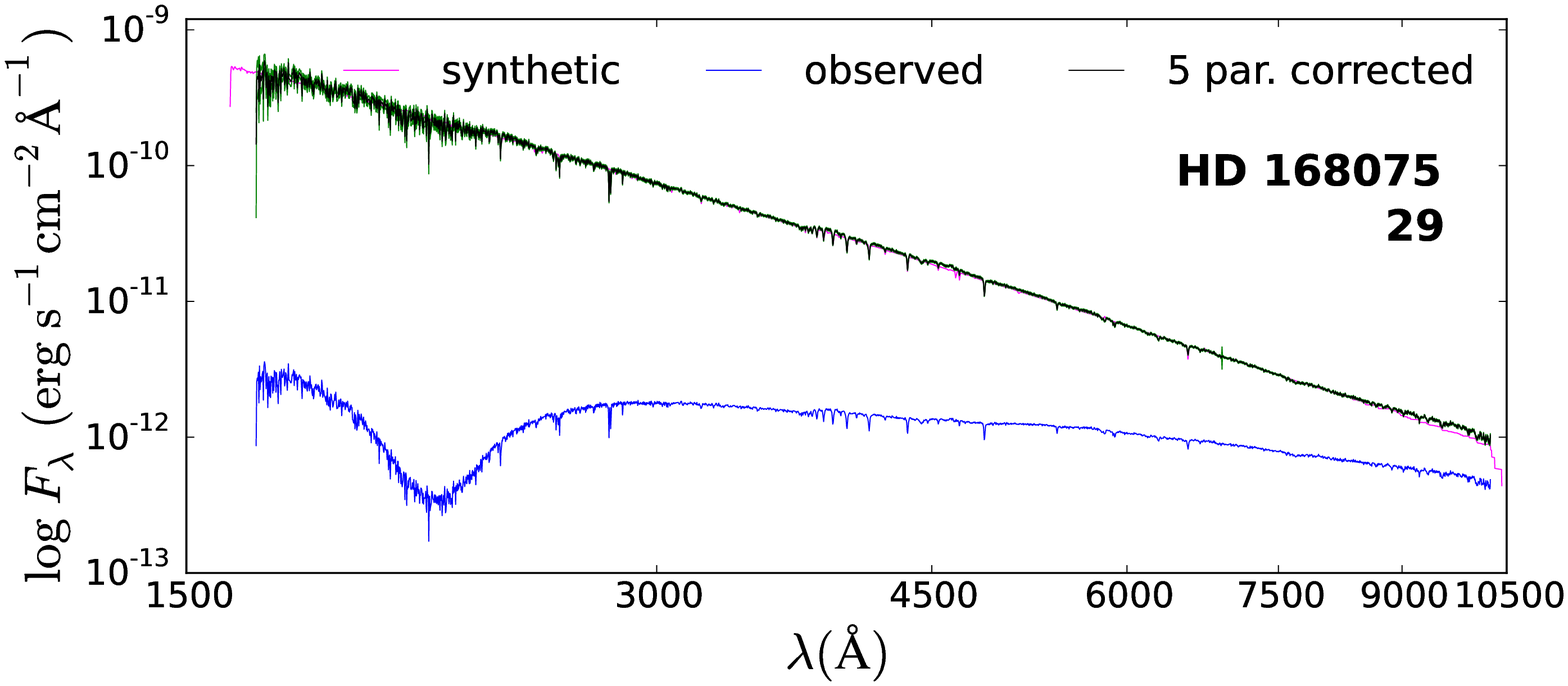}
\endminipage\hfill
\minipage{0.49\textwidth}
  \includegraphics[width=\linewidth]{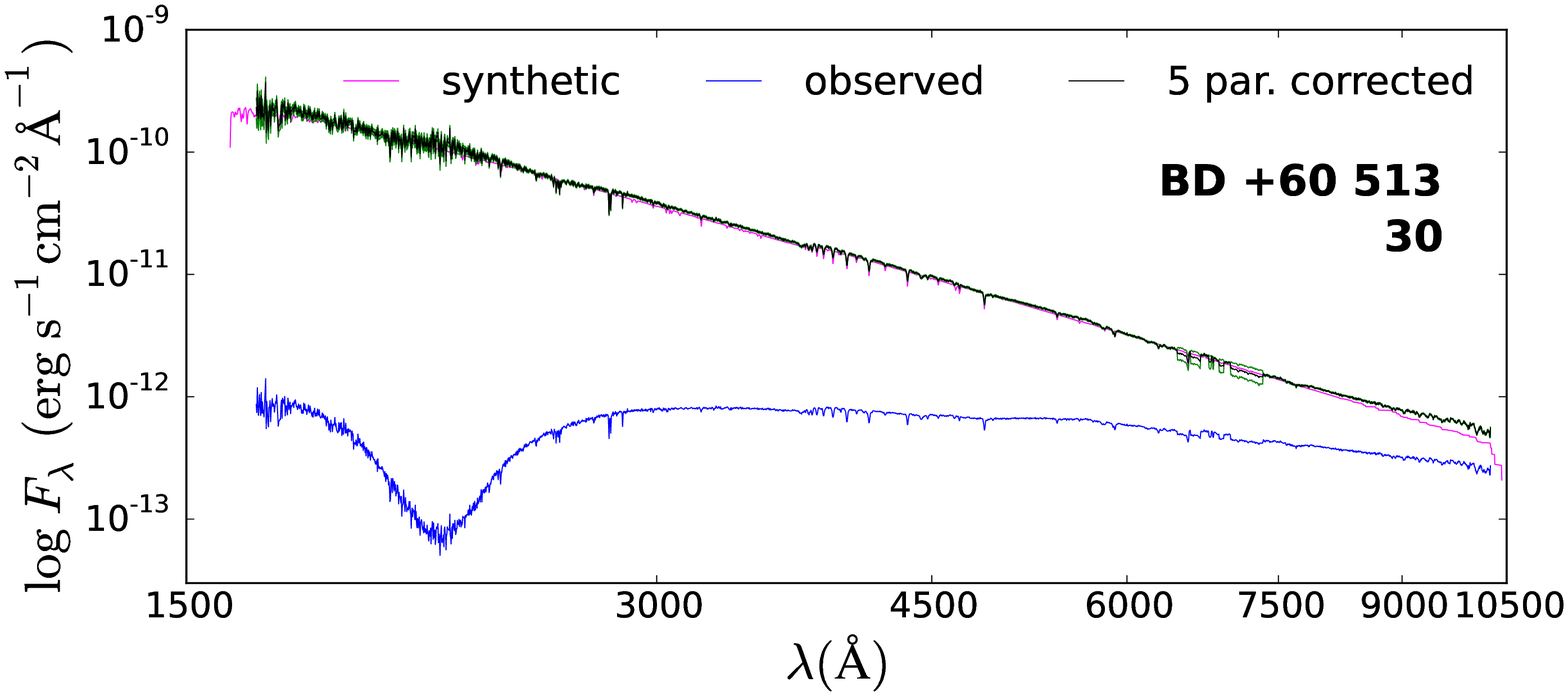} 
\endminipage\hfill
\minipage{0.49\textwidth}
  \includegraphics[width=\linewidth]{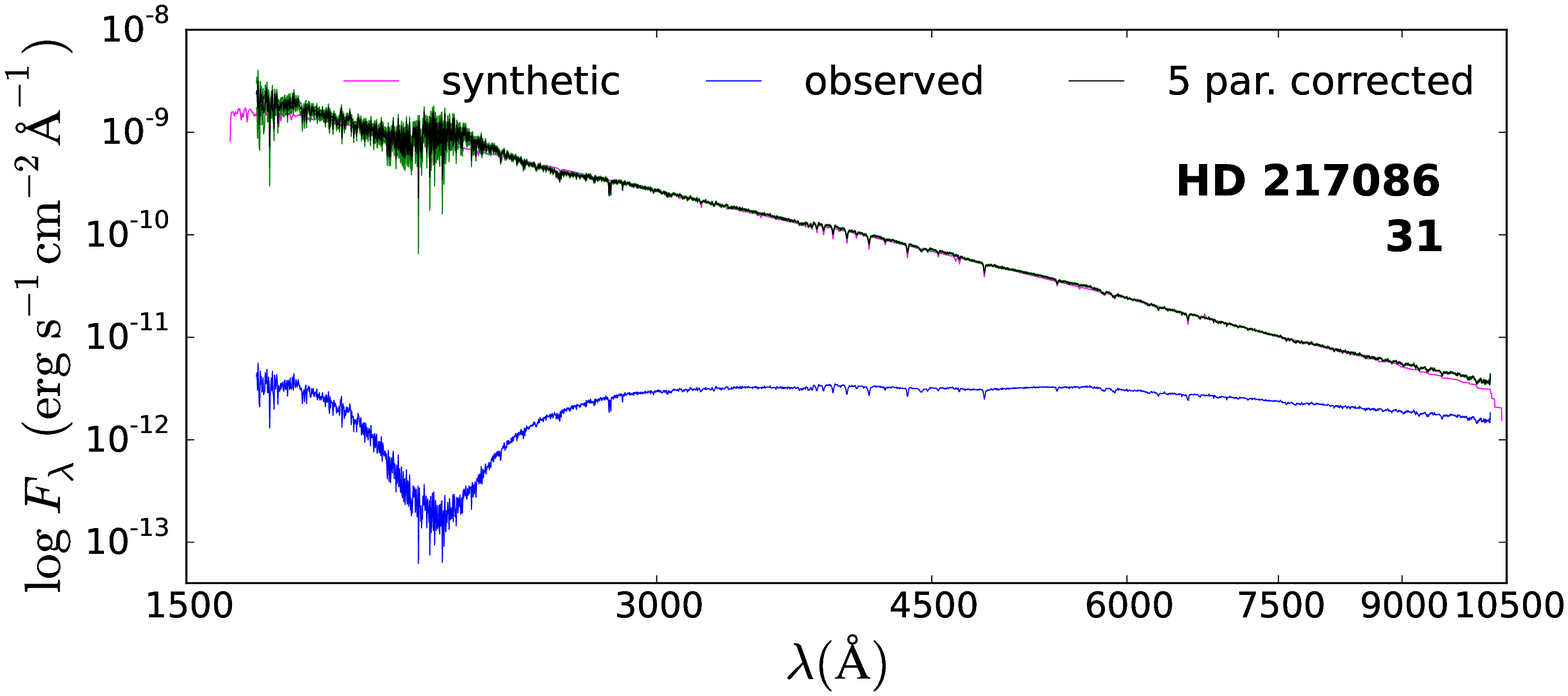}
\endminipage\hfill
\minipage{0.49\textwidth}
  \includegraphics[width=\linewidth]{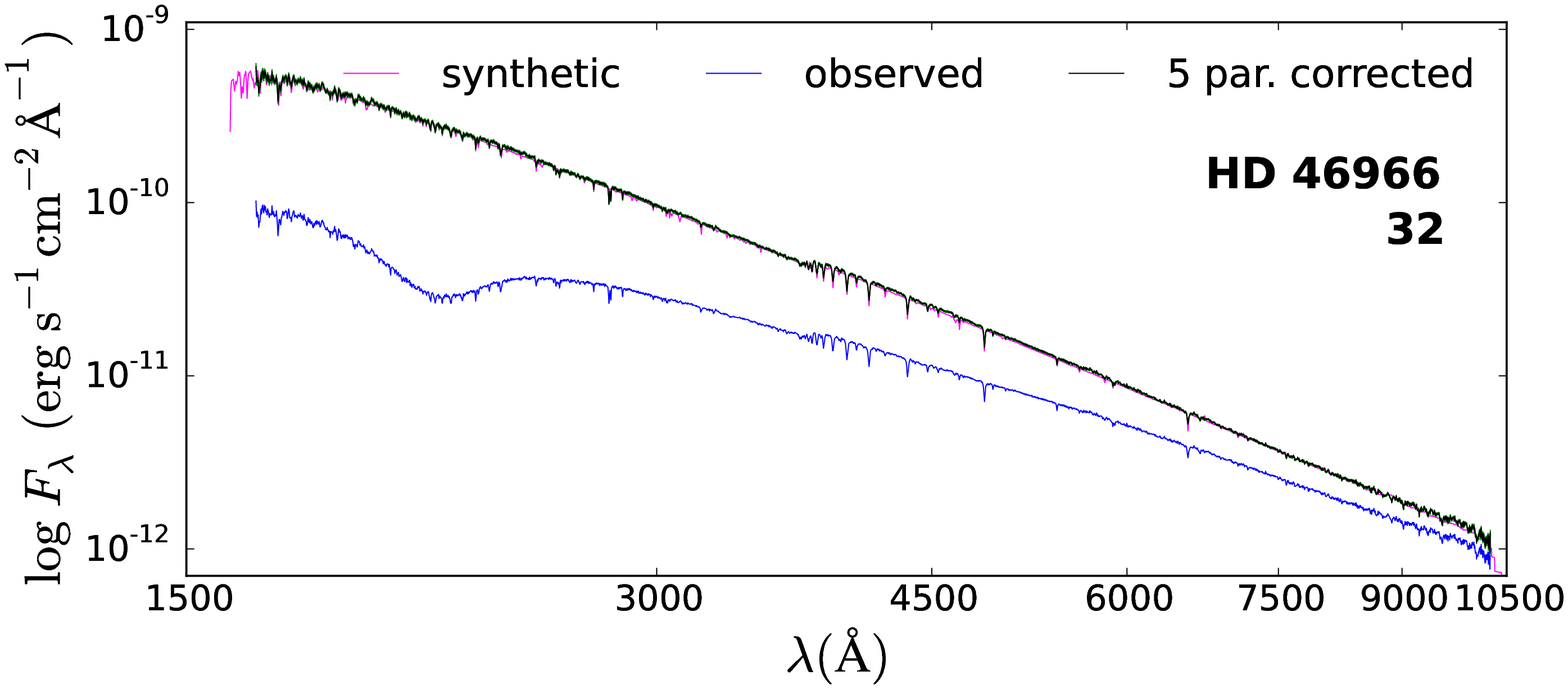}
\endminipage\hfill
\minipage{0.49\textwidth}
  \includegraphics[width=\linewidth]{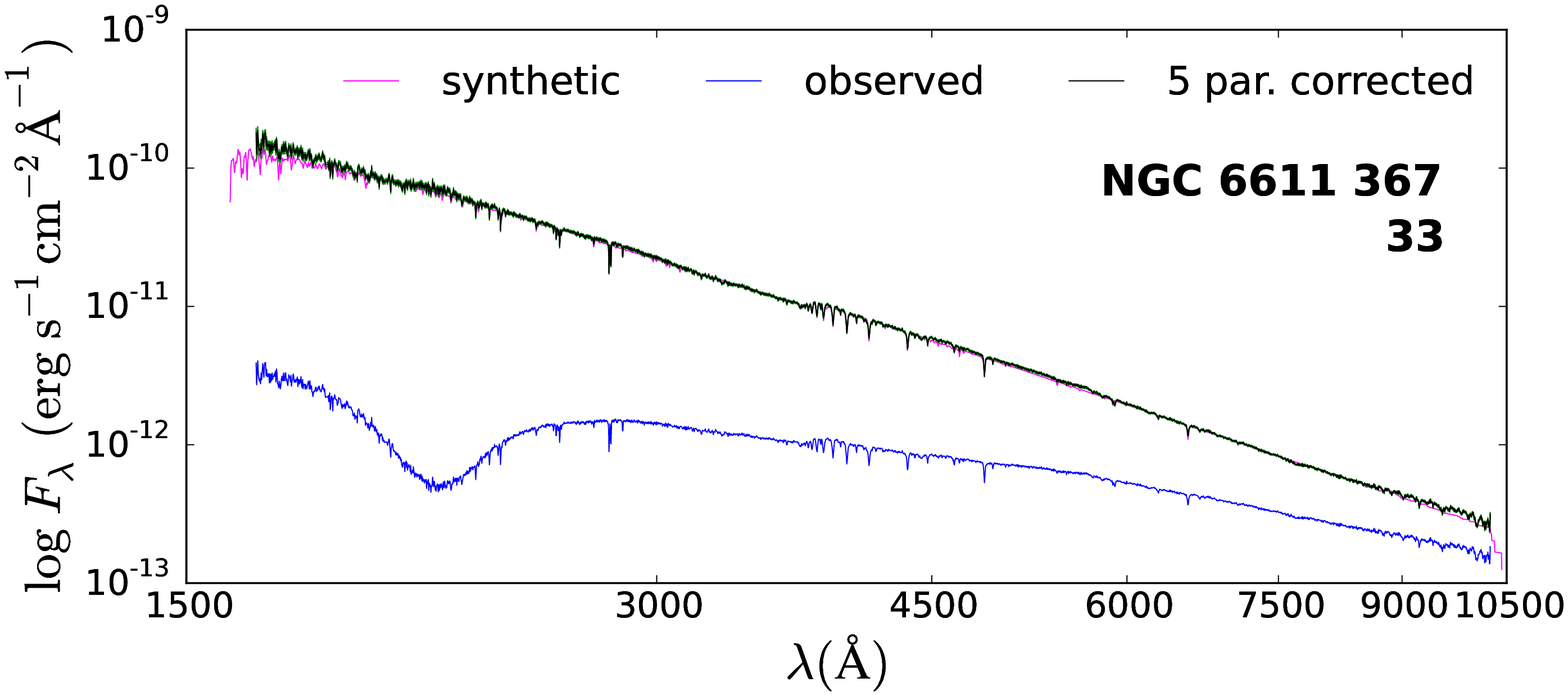}
\endminipage\hfill
\minipage{0.49\textwidth}
  \includegraphics[width=\linewidth]{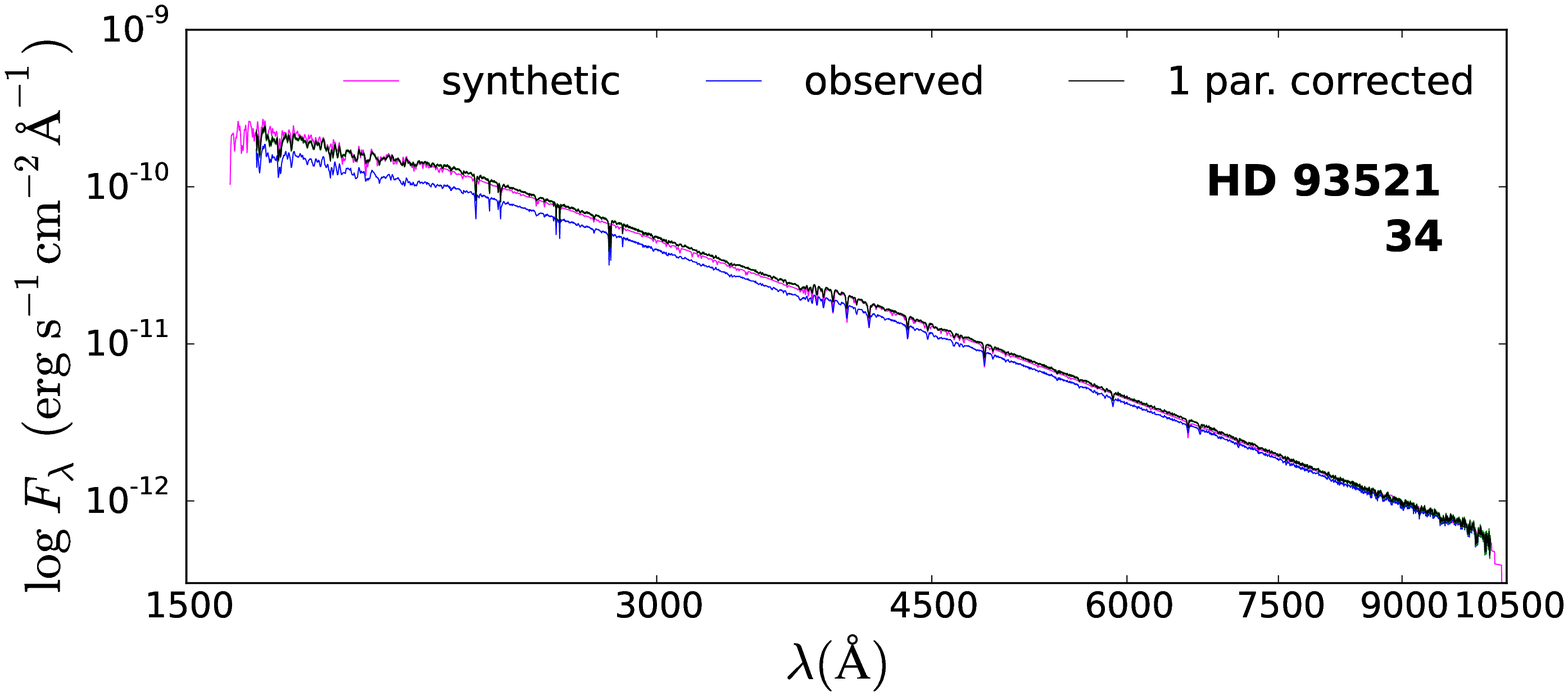}
\endminipage\hfill
\minipage{0.49\textwidth}
  \includegraphics[width=\linewidth]{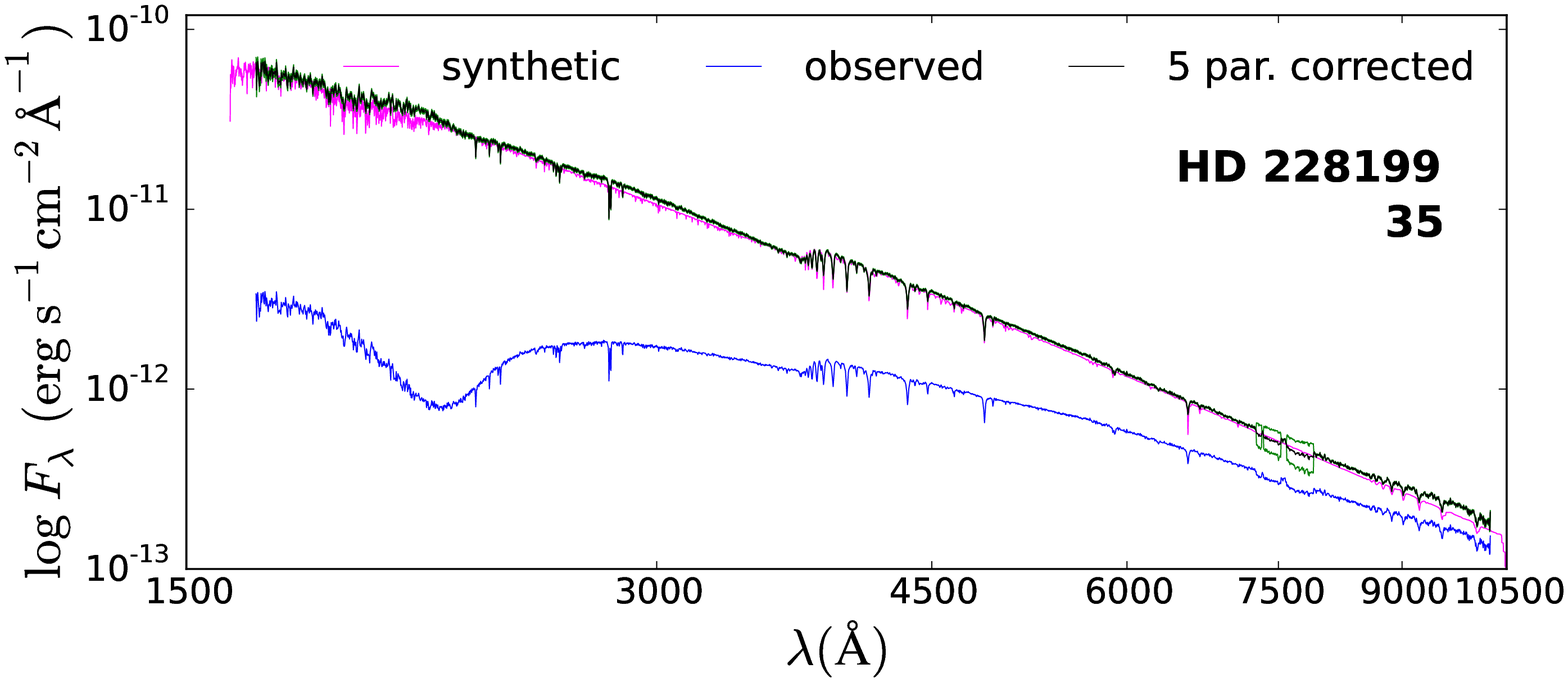}
\endminipage\hfill
\caption{Library spectra. Fluxes (blue) and extinction-corrected fluxes (black with
  green error bounds) are plotted along with the corresponding
  synthetic spectral template (magenta). }
\end{figure*}

\clearpage
\begin{figure*}[!htb]
\minipage{0.49\textwidth}
  \includegraphics[width=\linewidth]{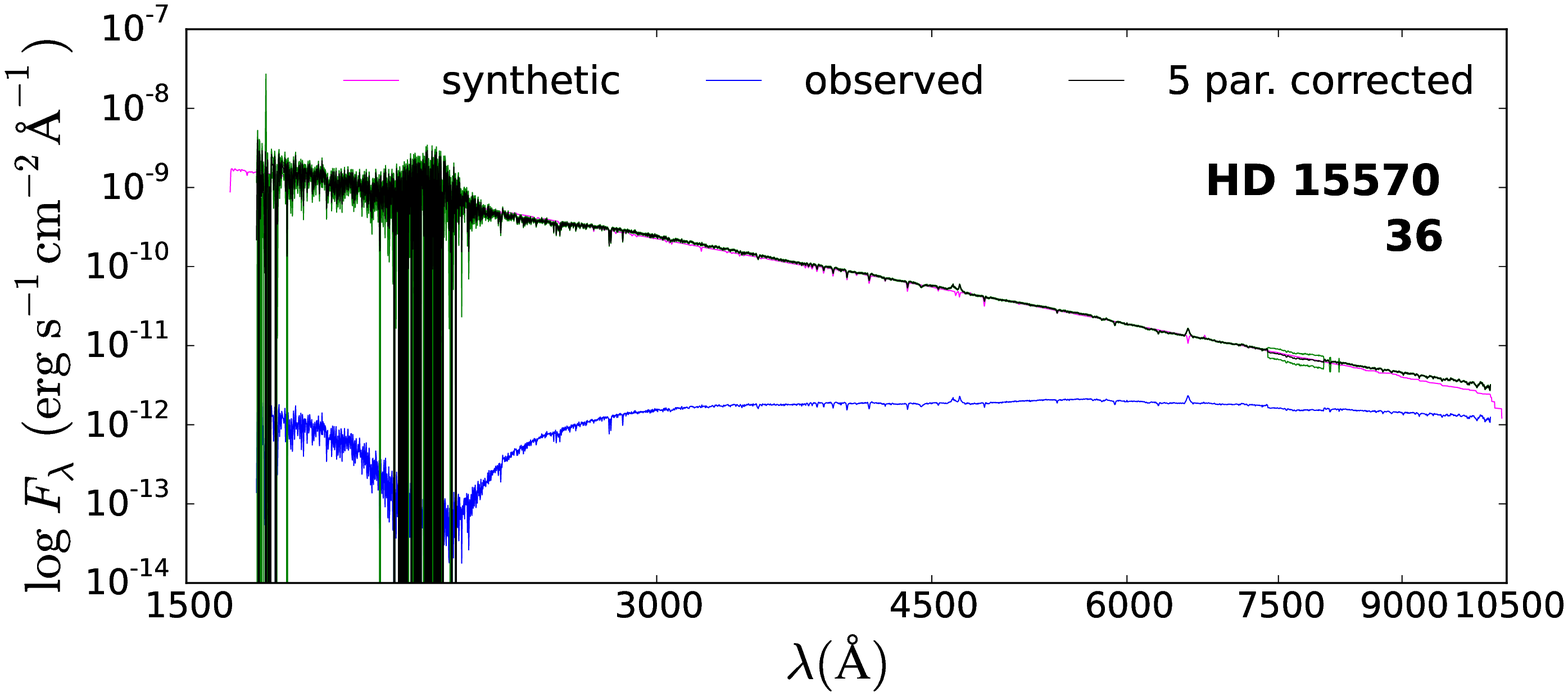}
\endminipage\hfill
\minipage{0.49\textwidth}
  \includegraphics[width=\linewidth]{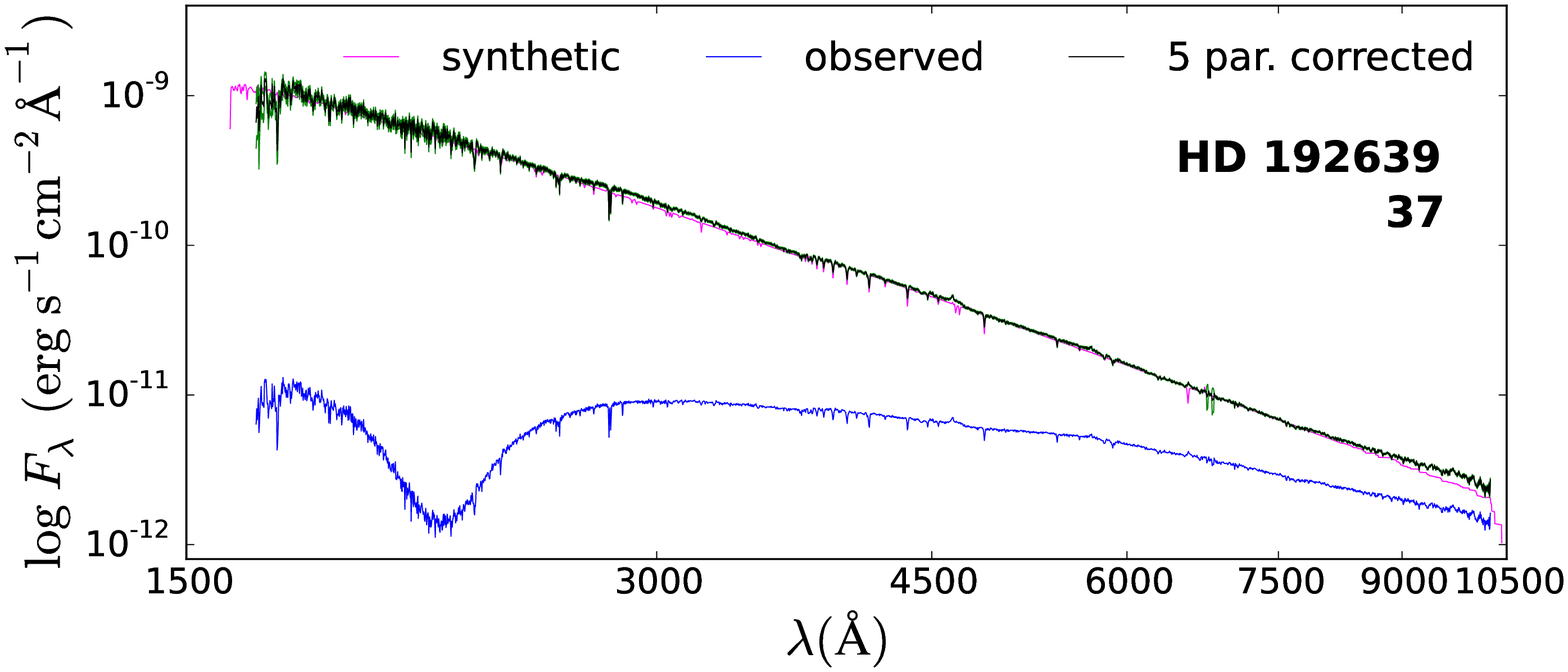} 
\endminipage\hfill
\minipage{0.49\textwidth}
  \includegraphics[width=\linewidth]{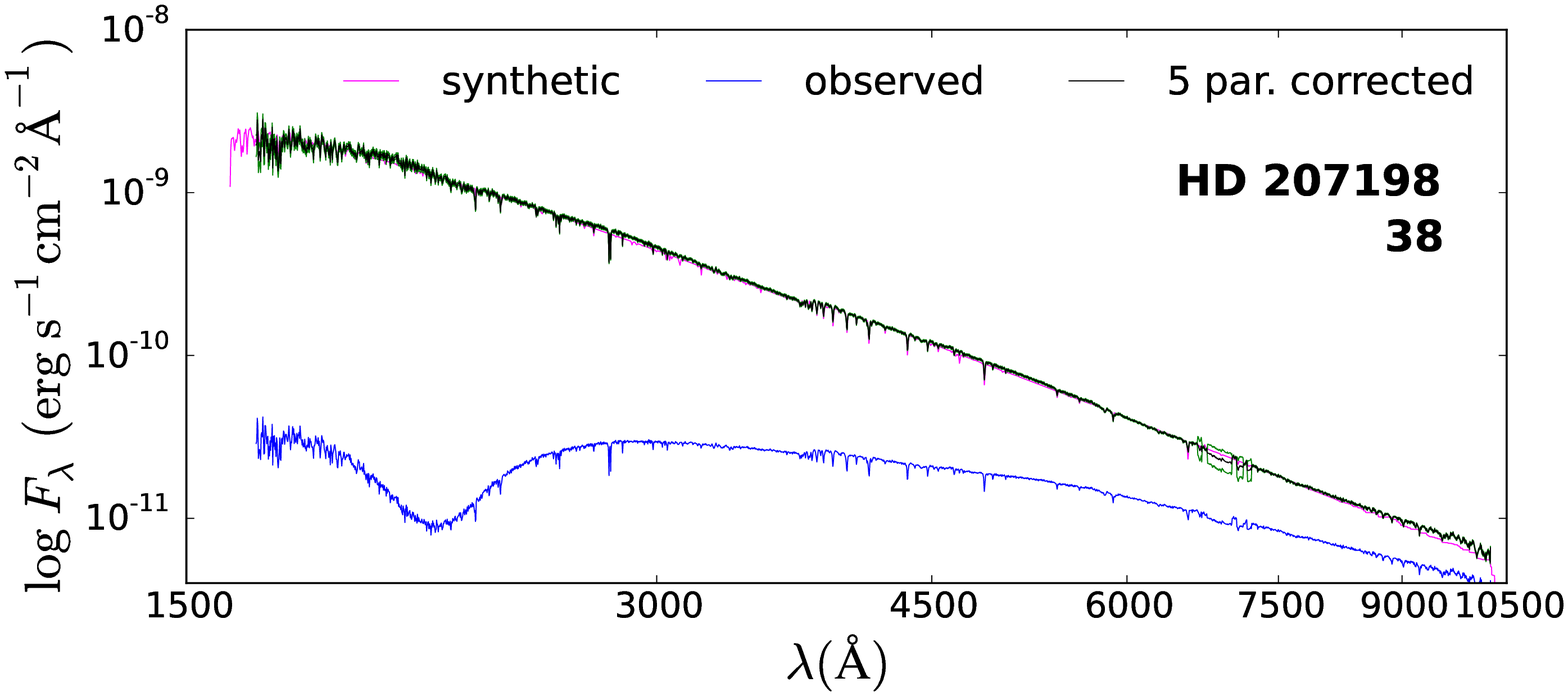}
\endminipage\hfill
\minipage{0.49\textwidth}
  \includegraphics[width=\linewidth]{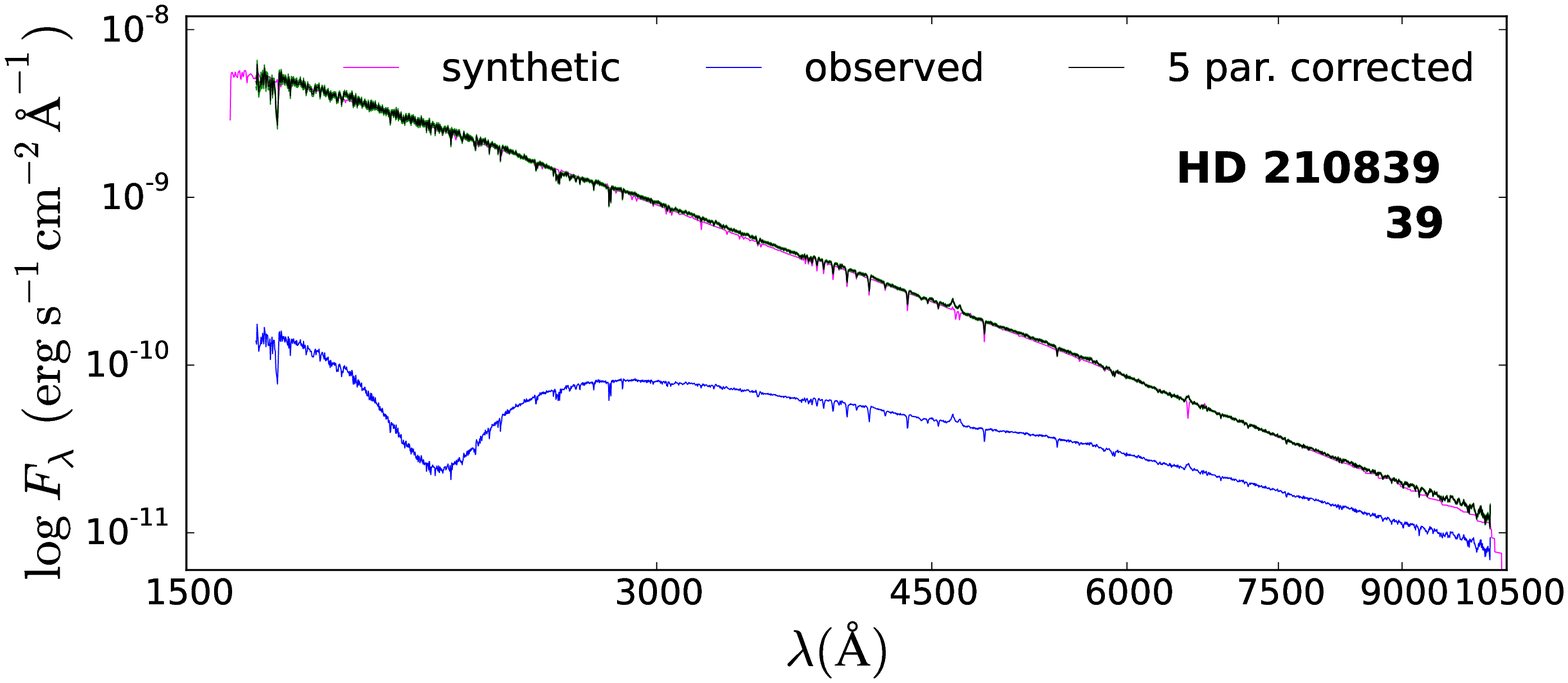}
\endminipage\hfill
\minipage{0.49\textwidth}
  \includegraphics[width=\linewidth]{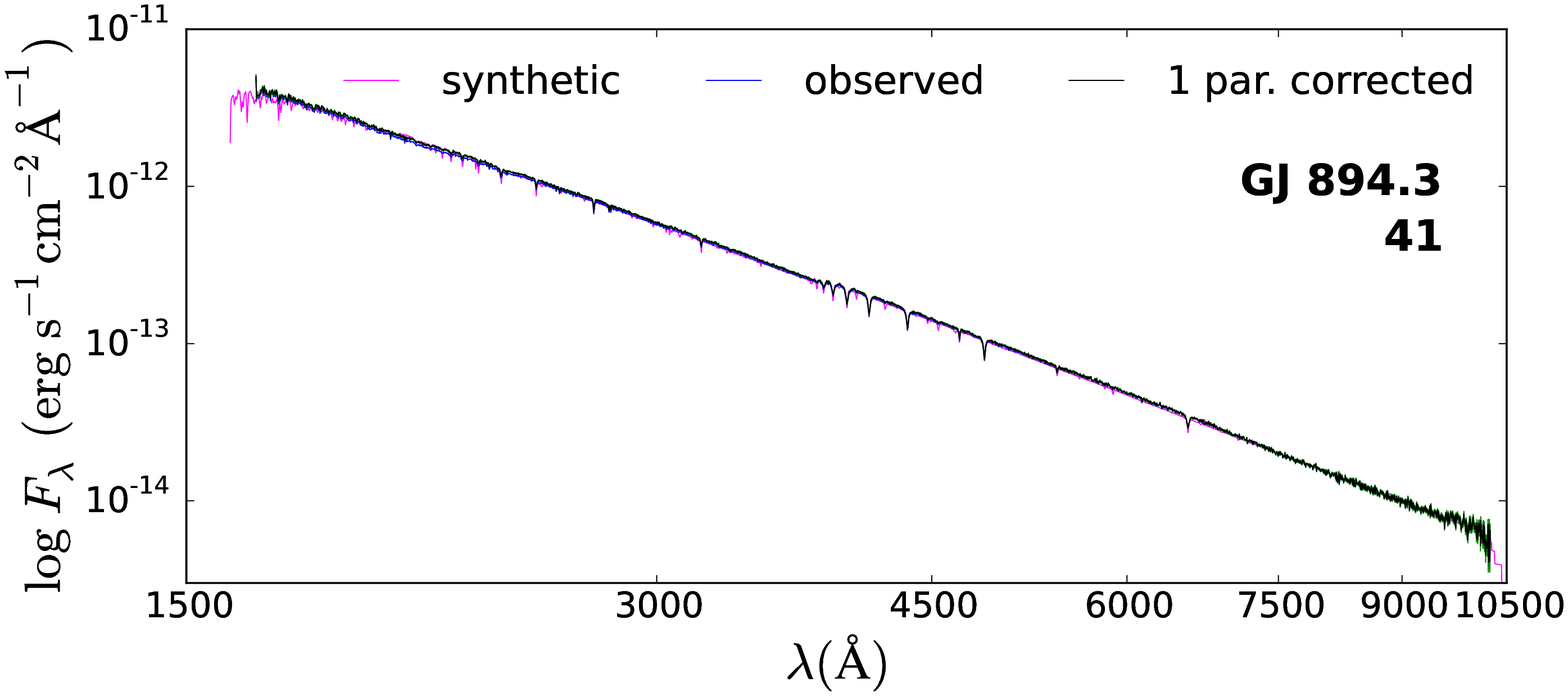}
\endminipage\hfill
\minipage{0.49\textwidth}
  \includegraphics[width=\linewidth]{fig_s43.eps} 
\endminipage\hfill
\minipage{0.49\textwidth}
  \includegraphics[width=\linewidth]{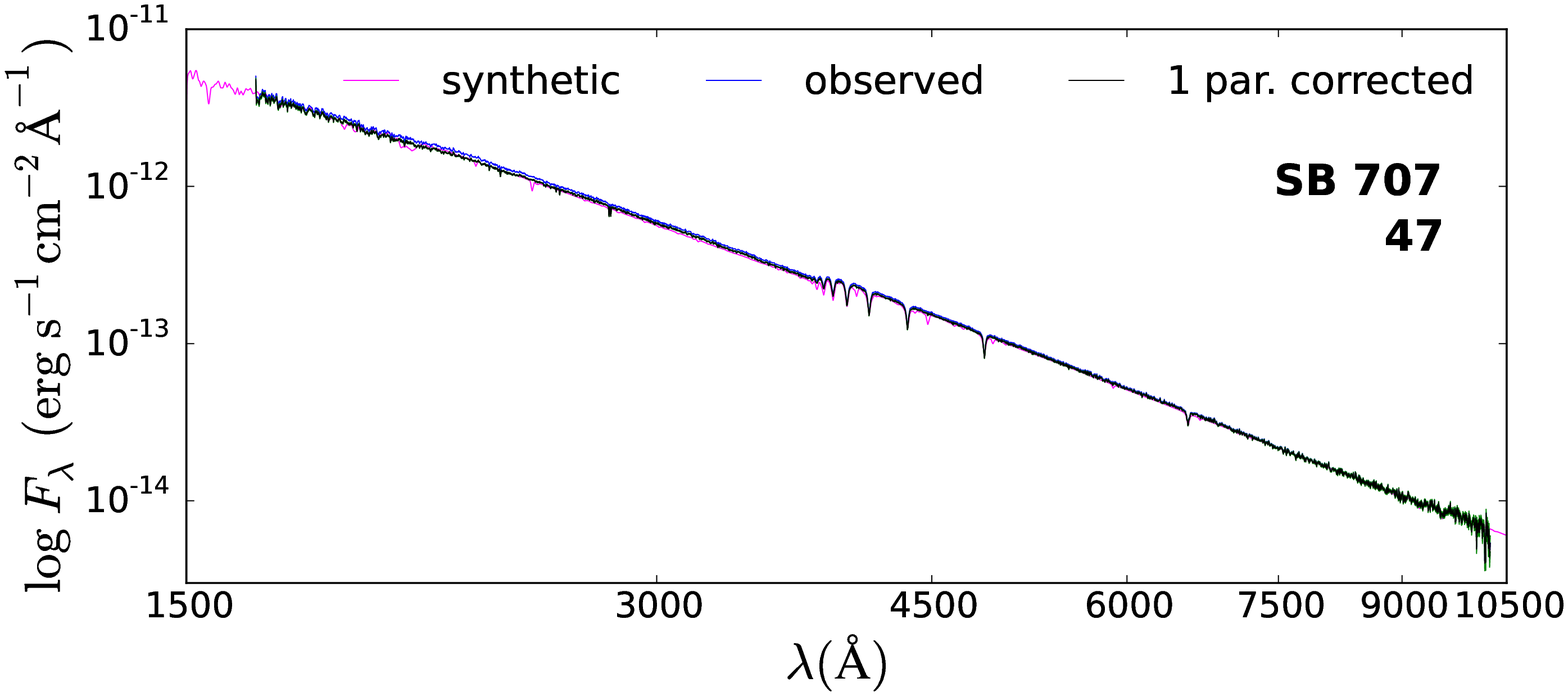}
\endminipage\hfill
\minipage{0.49\textwidth}
  \includegraphics[width=\linewidth]{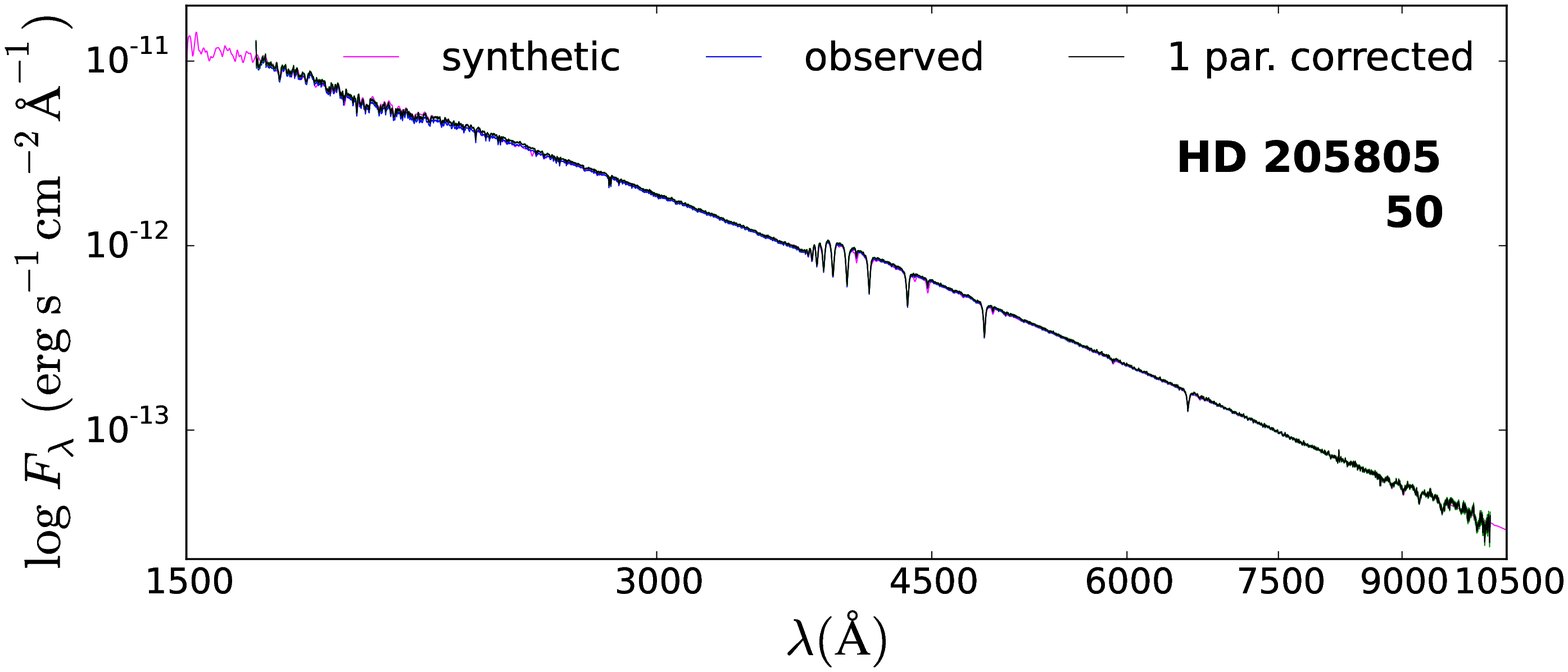}
\endminipage\hfill
\minipage{0.49\textwidth}
  \includegraphics[width=\linewidth]{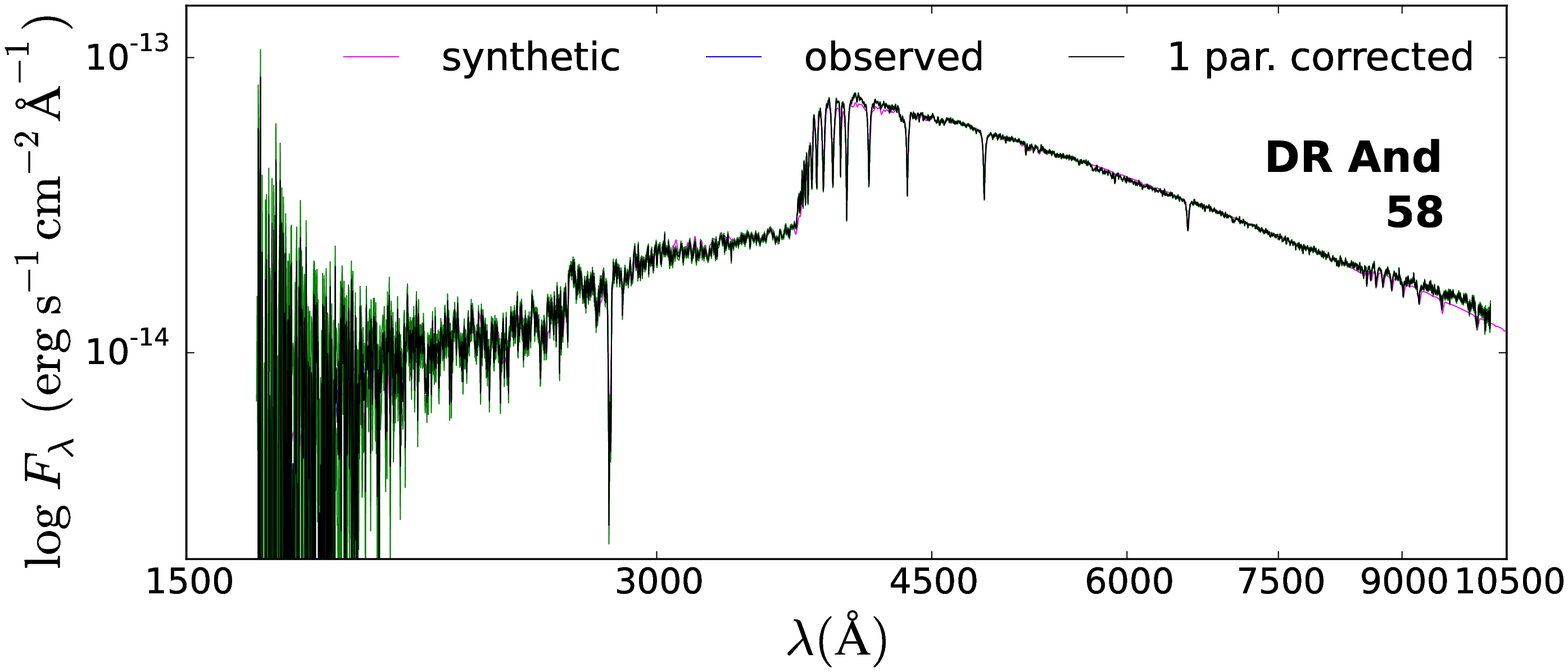} 
\endminipage\hfill
\minipage{0.49\textwidth}
  \includegraphics[width=\linewidth]{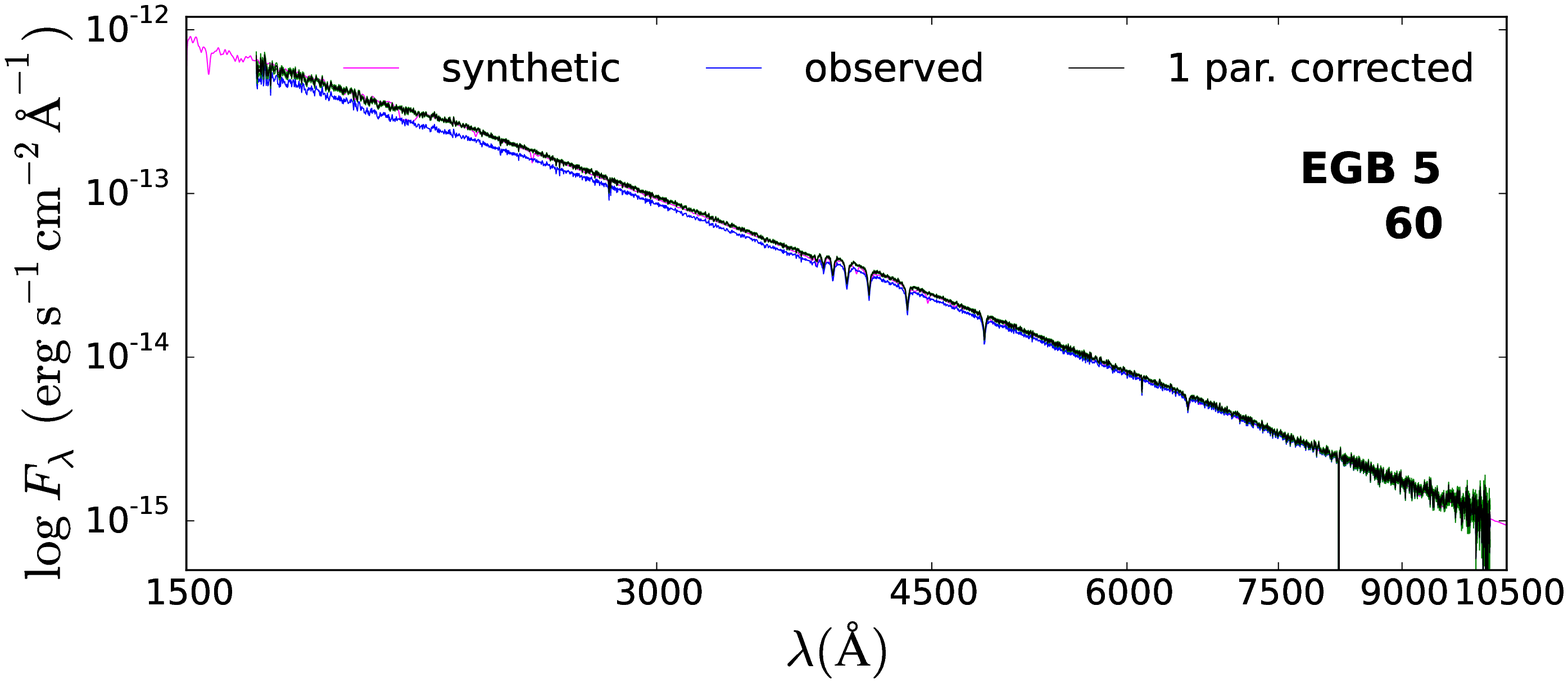}
\endminipage\hfill
\minipage{0.49\textwidth}
  \includegraphics[width=\linewidth]{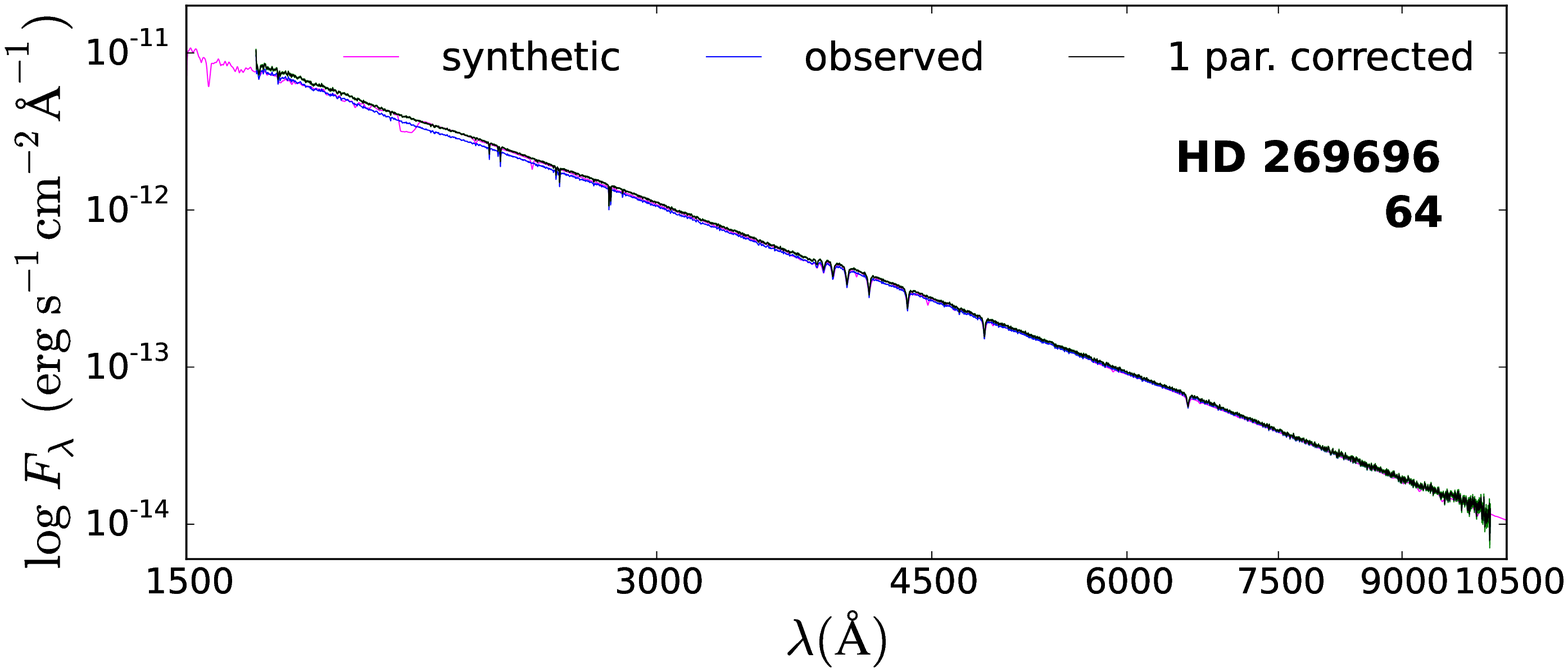}
\endminipage\hfill
\minipage{0.49\textwidth}
  \includegraphics[width=\linewidth]{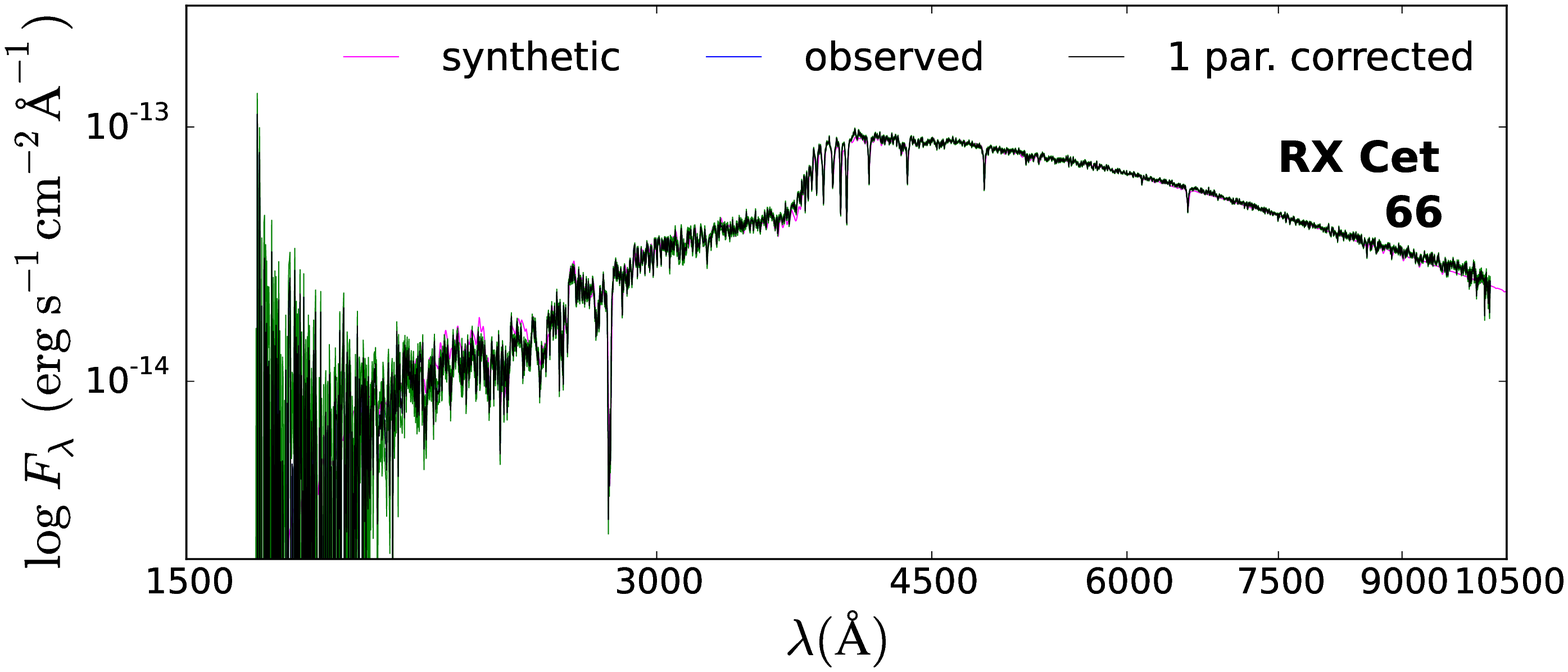}
\endminipage\hfill
\minipage{0.49\textwidth}
  \includegraphics[width=\linewidth]{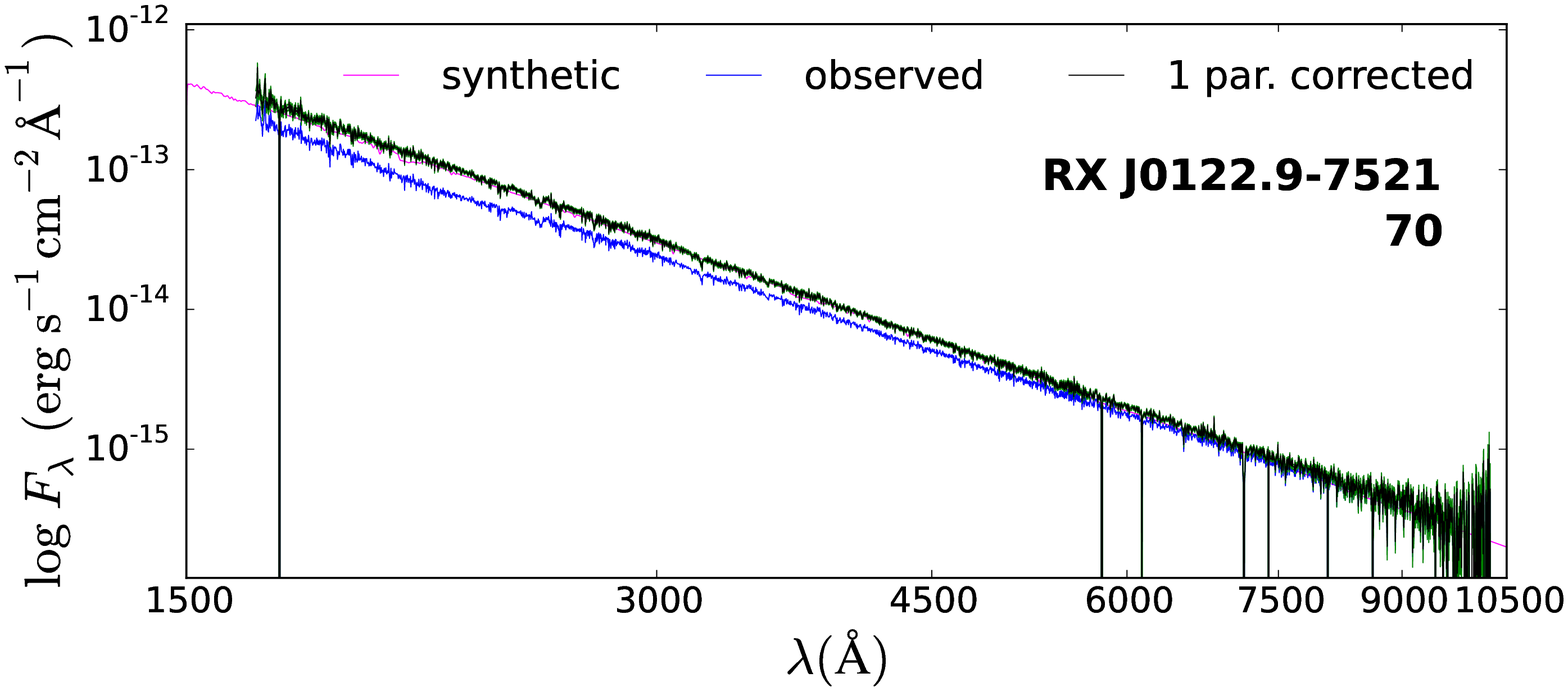}
\endminipage\hfill
\caption{Library spectra. Fluxes (blue) and extinction-corrected fluxes (black with
  green error bounds) are plotted along with the corresponding
  synthetic spectral template (magenta). }
\end{figure*}

\clearpage
\section{Dust extinction parameters}
Here we list 40 library stars with their dust extinction parameters. In Table \ref{table:1}, when only $A_V$ is listed, the other dust model parameters revert to their default values: $R = 3.1$ mag, $x_0 = 4.596\ \mu$m$^{-1}$, $\gamma = 0.99\ \mu$m$^{-1}$, $c_2 = -0.824 + 4.717 /  R$, $c_1 = 2.030 - 3.007 c_2$, and $c_3 = 3.23$.

\begin{table*}[!htb]
\caption{List of the 40 newly observed stars where the columns are as follows: internal catalog number based on the 70 SNAP-mode candidate stars, name of the star, apparent magnitude through the $V$ filter, color index, effective temperature, surface gravity, extinction in $V$, the ratio
of total to selective extinction at $V$, bump position, bump width, and bump strength. The internal catalog number becomes the HST dataset number if you prepend the characters ``ocy5,'' so that, for example, star 33 is HST dataset ocy533.}              
\label{table:1}     
\centering                                     
\resizebox{\textwidth}{!}{\begin{tabular}{l l c c c c c c c c c l}          
\hline\hline                       

N & Name & V & B-V & $T_{eff}$ & log($g$) & $A_v$ & $R$ & $x_0$  & $\gamma $ & $c_3$ & Reference \\
  &      & (mag) & (mag) & (K) &     & (mag) &    & $(\mu $m$^{-1})$  & $(\mu $m$^{-1})$ &   &  \\
\\
\hline      
&Post-AGB \\
\hline                             
    3 & NGC 1535   & 12.11 & $-0.038^*$& 70000 & 4.6 & 0.19 &&&&& \citet{2003IAUS..209..169W} \\ 
    4 & HD 107969  & 13.2  & $-0.4$ & 82000 & 5.5 & 0.07 &&&&& \citet{2003IAUS..209..169W} \\ 
    5 & LSS 1362   & 12.30 & $-0.22$& 100000 & 5.3 & 0.56 & 3.67 & 4.62 & 1.21 & 6.03 & \citet{2003IAUS..209..169W} \\ 
    6 & PN A66-78  & 13.25 & $-0.21$& 110000 & 5.5 & 0.57 & 3.24 & 4.70 & 1.07 & 2.88 & \citet{2003IAUS..209..169W} \\ 
    8 & NGC 7094   & 13.68 &$-0.050^*$& 110000 & 5.7 & 0.41 & 2.98 & 4.70 & 1.35 & 6.21 & \citet{2003IAUS..209..169W} \\ 
    9 & PN K 1-16  & 14.96 & $-0.52$& 140000 & 6.4 & 0.13 &&&&& \citet{2003IAUS..209..169W} \\ 
    10 & NGC 6853  & 14.09 & $-0.34$& 110000 & 6.7 & 0.08 &&&&& \citet{2003IAUS..209..169W} \\ 
    11 & LSV+46 21 & 12.87 & $-0.37$& 83000 & 6.7 & 0.16 &&&&& \citet{2003IAUS..209..169W} \\ 
    13 & LSIII+52 24 & 12.51 & 0.72 & 24000 & 3.0 & 2.07 &  3.08 &  4.89 &  2.85 &  40.64 & \citet{Sarker2012}\\ 
    14 & LSIV-12 111 & 11.4 & $-0.1$& 20500 & 2.35 & 0.88 & 2.98 & 4.63 & 1.16 & 4.14 & \citet{Ryans2003} \\ 
    16 & GLMP 334  & 12.3  & 0.8    & 7400 & 1.4 & 1.81 &  2.79 & 4.60 &  0.99  &  3.50 & \citet{refId0}   \\ 
    17 & NGC 3242  & 10.3  & 0.0    & 68000 & 4.6 & 0.20 & 2.8 & 4.81 & 1.27 & 4.57 & \citet{Mendez1988}\\ 
    18 & NGC 1535  & 12.11 &$-0.038^*$& 58000 & 4.3 & 0.21 & 2.83 & 4.73 & 1.13 & 3.65 & \citet{Mendez1988}\\ 
    19 & IC 2448   & 11.1  & 0.0    & 55000 & 4.5 & 0.20 &&&&& \citet{Mendez1988} \\
    21 & NGC 2392  & 10.52  &$-0.15$ & 47000 & 3.6 & 0.44 & 3.27 & 4.66 & 1.11 & 1.82 & \citet{Mendez1988}\\ 
    22 & Hen 2-182 & 11.72 & 0.31   & 36000 & 3.4 & 0.86 & 3.75 & 4.76 & 2.12 & 20.4 & \citet{Mendez1988}\\
    23 & PN Tc 1   & 11.49 & $-0.19$& 33000 & 3.2 & 0.62 & 3.06 & 4.67 & 1.03 & 3.34 & \citet{Mendez1988}\\
\\
\hline
&Normal O Type \\  
\hline                         
    25 & HD 93250 & 7.50 & 0.16 & 51000 & 3.90 & 1.70 & 3.49 & 4.64 & 1.05 &  3.24 & \citet{1996ApJ...460..914V} \\ 
    26 & HDE 303308 & 6.17 & 0.13 & 48000 & 3.91 & 1.23 & 3.00 & 4.60 & 0.86 & 2.52 & \citet{1996ApJ...460..914V} \\ 
    27 & HD 164794 & 5.97 & 0.0 & 47000 & 3.90 & 1.02 & 3.28 & 4.58 & 0.94 & 3.26 & \citet{1996ApJ...460..914V}\\ 
    28 & HD 46150 & 6.73 & 0.13 & 46900 & 3.85 & 1.08 & 2.82 & 4.61 & 0.91 & 3.57 & \citet{1996ApJ...460..914V} \\ 
    29 & HD 168075 & 8.77 & 0.55 & 49000 & 3.95 & 2.28 &  3.15 &  4.62 & 0.99 &  3.23 & \citet{1996ApJ...460..914V}\\ 
    30 & BD +60 513 & 9.39 & 0.49 & 40000 & 3.70 & 2.17 &  2.85 &  4.58 &  0.89 & 3.32 & \citet{1996ApJ...460..914V}\\ 
    31 & HD 217086 & 7.66 & 0.64 & 40000 & 3.71 & 2.62 & 2.96 & 4.57 & 0.81 & 2.83 & \citet{1996ApJ...460..914V}\\ 
    32 & HD 46966 & 6.87 & $-0.04$ & 38100 & 3.89 & 0.66 & 2.77 & 4.62 & 1.04 & 4.42 & \citet{1996ApJ...460..914V}\\ 
    33 & NGC 6611 367 & 9.45 & 0.26 & 35000 & 4.15 & 1.65 & 3.07 & 4.61 & 0.85 & 2.63 & \citet{1996ApJ...460..914V}\\
    34 & HD 93521 & 7.03 & $-0.24$ & 33500 & 3.95 & 0.11 &&&&& \citet{1996ApJ...460..914V}\\ 
    35 & HD 228199 & 9.34 & 0.07 & 30000 & 3.91 & 0.96 & 2.53 & 4.64 & 0.92 &  3.50 & \citet{1996ApJ...460..914V}\\ 
    36 & HD 15570 & 8.11 & 0.69 & 49000 & 3.51 & 2.85 & 2.87 & 4.62 & 0.72 & 2.30 & \citet{1996ApJ...460..914V}\\ 
    37 & HD 192639 & 7.11 & 0.35 & 38000 & 3.37 & 1.59 & 2.62 & 4.58 & 0.86 & 3.15 & \citet{1996ApJ...460..914V}\\ 
    38 & HD 207198 & 5.94 & 0.31 & 34000 & 3.31 & 1.44 & 2.63 & 4.65 & 1.08 & 3.96 & \citet{1996ApJ...460..914V}\\ 
    39 & HD 210839 & 5.05 & 0.24 & 40000 & 3.50 & 1.36 & 2.81 & 4.60 & 1.00 &  4.31 & \citet{1996ApJ...460..914V}\\ \\
\hline
&Horizontal Branch \\  
\hline      
    41 & GJ 894.3 & 11.50 & $-0.05$ & 40000 & 5.00 & 0.01 & & & & & \citet{Heber1984} \\ 
    43 & PHL 382 & 11.4 & $-0.1$ & 18200 & 4.1 & 0.10 & & & & & \citet{1987MitAG..70...79H} \\ 
    47 & SB 707 & 11.9 & $-0.3$ & 34000 & 6.0 & 0.00 & & & & & \citet{1985ApJ...299..496L}\\ 
    50 & HD 205805 & 10.158 & $-0.241$ & 25000 & 5.0 & 0.01 &&&&& \citet{Altmann2000}\\ 
    58 & DR And & 12.42 & 1.24 & 6200 & 2.3 & 0.00 & & & & & \citet{2015MNRAS.447.2404P} \\ 
    60 & EGB 5 & 14.2 & $-0.9$ & 42000 & 5.8 & 0.06 & & & & & \citet{Mendez1988}\\ 
    64 & HD 269696 & 11.138 & $-0.297$ & 40000 & 5.5 & 0.03 & & & & & \citet{1984MNRAS.209..387L}\\ 
    66 & RX Cet & 11.2 & 0.4 & 6186 & 2.0 & 0.00 & & & & & \citet{2015MNRAS.447.2404P} \\ 
    70 & RX J0122.9-7521 & 15.4 & $-0.4$ & 180000 & 7.5 & 0.54 & 4.54 & 4.53 & 1.20 & 2.02 & \citet{2006PASP..118..183W} \\
     \\
                                       
\end{tabular}}
$^*$PAGB stars missing $B-V$ have, instead, $J-K$ from the 2MASS point source catalog.
\end{table*}

\end{document}